\documentclass[10pt,twocolumn,showpacs,longbibliography,amsmath,amssymb,osajnl,floatfix,superscriptaddress]{revtex4-1}

\usepackage{graphicx}
\usepackage{dcolumn}
\usepackage{bm}
\usepackage{color}
\usepackage{txfonts}
\usepackage{microtype}
\usepackage{float}

\begin{document}

\author{Kun Xue}
 \affiliation{MOE Key Laboratory for Nonequilibrium Synthesis and Modulation of Condensed Matter, School of Physics, Xi'an Jiaotong University, Xi'an 710049, China}

\author{Ren-Tong Guo}
 \affiliation{MOE Key Laboratory for Nonequilibrium Synthesis and Modulation of Condensed Matter, School of Physics, Xi'an Jiaotong University, Xi'an 710049, China}

\author{Feng Wan}	
\affiliation{MOE Key Laboratory for Nonequilibrium Synthesis and Modulation of Condensed Matter, School of Physics, Xi'an Jiaotong University, Xi'an 710049, China}	

\author{Rashid Shaisultanov} \affiliation{Max-Planck-Institut f\"{u}r Kernphysik, Saupfercheckweg 1,
	69117 Heidelberg, Germany} \affiliation{Helmholtz-Zentrum Dresden-Rossendorf, Bautzner Landstra{\ss}e 400, 01328 
Dresden, Germany}

\author{Yue-Yue Chen}\affiliation{Department of Physics, Shanghai Normal University, Shanghai 200234, China}

\author{Zhong-Feng Xu}  	\affiliation{MOE Key Laboratory for Nonequilibrium Synthesis and Modulation of Condensed Matter, School of Physics, Xi'an Jiaotong University, Xi'an 710049, China}

\author{Xue-Guang Ren}  	\affiliation{MOE Key Laboratory for Nonequilibrium Synthesis and Modulation of Condensed Matter, School of Physics, Xi'an Jiaotong University, Xi'an 710049, China}	
		
\author{Karen Z. Hatsagortsyan}\email{k.hatsagortsyan@mpi-hd.mpg.de}
\affiliation{Max-Planck-Institut f\"{u}r Kernphysik, Saupfercheckweg 1,	69117 Heidelberg, Germany}

\author{Christoph H. Keitel}
\affiliation{Max-Planck-Institut f\"{u}r Kernphysik, Saupfercheckweg 1,
	69117 Heidelberg, Germany}
	
\author{Jian-Xing Li}\email{jianxing@xjtu.edu.cn}
\affiliation{MOE Key Laboratory for Nonequilibrium Synthesis and Modulation of Condensed Matter, School of Physics, Xi'an Jiaotong University, Xi'an 710049, China}

\title{Generation of arbitrarily polarized GeV lepton beams via nonlinear Breit-Wheeler process}

\date{\today}

\begin{abstract}

Generation of arbitrarily spin-polarized lepton
(here refer in particular to electron and positron) beams has been investigated in the single-shot interaction of high-energy polarized $\gamma$  photons with an ultraintense asymmetric laser pulse via nonlinear Breit-Wheeler (BW) pair production.
We develop a fully spin-resolved semi-classical Monte Carlo method to describe the pair creation and polarization in the local constant field approximation. 
In nonlinear BW process the polarization of created pairs is simultaneously determined by the polarization of parent $\gamma$ photons, the polarization and asymmetry of scattering laser field, due to the spin angular momentum transfer and the asymmetric spin-dependent pair production probabilities, respectively.
In considered all-optical method, dense GeV lepton beams with average polarization degree up to about $80\%$ (adjustable between the transverse and longitudinal components) can be obtained with currently achievable laser facilities, which could be used as injectors of the polarized $e^{+}e^{-}$  collider to search for new physics beyond the Standard Model.

\end{abstract}

\maketitle
Ultrarelativistic spin-polarized lepton (here refer in particular to electron and positron) beams have many important applications in particle and high-energy physics \cite{Wardle_1998,RevModPhys.76.323,Ablikim2018zay}, especially in $e^{+}e^{-}$ collider, such as  International Linear Collider (ILC) \cite{Moortgat2008,baer2013},  Compact Linear Collider (CLIC) \cite{Simitcioglu2018,Ari2016} and  Circular Electron Positron Collider (CEPC) \cite{Duan2019,Nikitin2020}. In those experiments, the longitudinal polarization of leptons can change interaction cross section and consequently  provides high sensitivities \cite{Moortgat2008} through, e.g., suppressing background from $WW$ boson and single $Z$ boson production via $WW$ fusion \cite{Moortgat2008}, enhancing different triple gauge couplings in $WW$ pair production 
\cite{Moortgat2008,Diehl2003} and improving top vector coupling in top quark production \cite{chakraborty2003}; while, the transverse polarization can cause asymmetric azimuthal distribution of  final-state particles \cite{Bartl2007} and then brings a way to study new physics beyond the Standard Model (BSM) \cite{Herczeg2003, Ananthanarayan2004, Ananthanarayan_2005,Ananthanarayan2018} through, e.g., measuring relative phases among
helicity amplitudes in $WW$ pair production \cite{Fleischer1994}, probing mixture of scalar-electron states \cite{Hikasa_1986} and searching for graviton in extra dimensions \cite{Rizzo2003}. Commonly, longitudinal and transverse polarizations are studied separately since those corresponding effects are independent
to each other \cite{Moortgat2008,Bartl2007}. However, it deserves to point out that the arbitrarily spin-polarized (ASP) lepton beams (having both longitudinal and transverse polarization components) also attract broad interests, because they can introduce three mutually orthogonal axes (required by fully reconstructing the density matrix of a spin-1/2 particle), modify effective BSM vertices \cite{Dass1977,Burgess1991,Ananthanarayan2004}, and thus play an unique role in future BSM experiments in $e^{+}e^{-}$ colliders, e.g., rendering special spin structure functions as being observable in vector and axial-vector type BSM interactions \cite{Ananthanarayan2018}, producing polarized top quark pairs as a probe of new physics \cite{Harlander1997,Godbole_2006} and diagnosing spin and chirality structures of new particles in antler-topology processes \cite{Choi2015}.

In conventional methods, the transversely polarized lepton beams can be directly obtained in a storage ring via Sokolov-Ternov effect \cite{Sokolov_1964,Sokolov_1968,Baier_1967,Baier_1972,Derbenev_1973}, which demands a long polarization time since large-size static magnetic fields are relatively weak ($\sim$Tesla), and the longitudinally polarized ones can be created
via high-energy circularly polarized (CP) $\gamma$ photons interacting with high-$Z$ target \cite{Omori_2006,Alexander_2008,abbott2016prl} (Bethe-Heitler $e^+e^-$ pair production process \cite{Heitler_1954}), in which the low luminosity of $\gamma$ photons  
requires a large amount of repetitions to yield a dense positron beam for further applications \cite{Variola_2014}.
 The transverse and longitudinal polarizations can be transformed to each other by a spin rotator, which demands quasi-monoenergetic beams with a risk of beam intensity reduction \cite{Buon_1986,Moffeit_2005}.

Modern ultrashort ultraintense laser pulses 
\cite{Yoon2019,Danson_2019,Gales_2018} enable alternative efficient methods to generate dense polarized lepton beams in femtoseconds via nonlinear quantum electrodynamics (QED) processes \cite{Ritus_1985}, e.g. nonlinear Compton scattering \cite{Goldman_1964, Nikishov_1964, Brown_1964,CAIN} and Breit-Wheeler (BW) $e^+e^-$ pair production \cite{Reiss1962,Ivanov_2005, Seipt_2020,Wistisen_2020,Wan_2020}. 
As reported, the leptons can be greatly transversely polarized in a standing-wave   \cite{Sorbo_2017,Sorbo_2018,Seipt_2018}, elliptically polarized (EP) \cite{li2019prl}, or bichromatic laser \cite{Seipt_2019,Song_2019,Chen_2019}  but not in a monochromatic symmetric laser \cite{Kotkin2003prstab,Ivanov_2004,Karlovets_2011}. And, longitudinally polarized positrons can be produced by CP $\gamma$ photons through the helicity transfer~\cite{Li_2020_2} (similar to the Bethe-Heitler process). Moreover, two-step schemes have also been proposed: low-energy polarized leptons are first generated by polarized photocathodes \cite{Pierce1980,Kuwahara2012,Zitzewitz1979}, polarized atoms \cite{Barth2013} or molecular photodissociation \cite{Rakitzis1936,sofikitis2017,sofikitis2018}, and then accelerated to ultrarelativistic energies via laser- \cite{Wen_2019,Yitong2019b} or beam-driven \cite{Yitong2019a,Nie2021} plasma wakefield (conventional accelerators work as well). All those proposals provide leptons  either greatly transversely or longitudinally polarized, however, how to generate
above mentioned unique ASP lepton beams is still a great challenge. 
 
 \begin{figure}[hptb]
	\setlength{\abovecaptionskip}{-0.3cm}
	\includegraphics[width=1.0\linewidth]{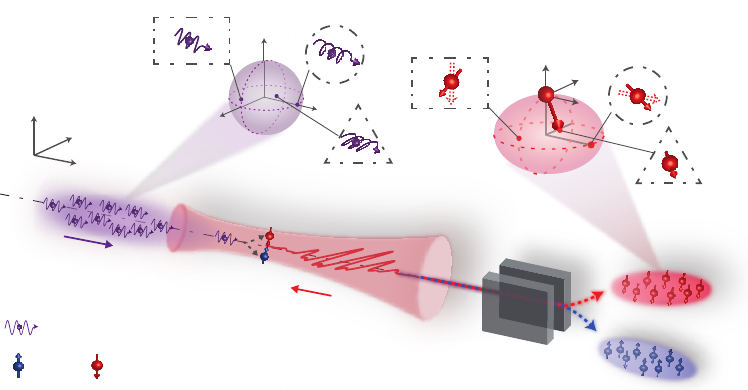}
	\begin{picture}(300,0)
	\put(10,104){\small $\hat{y}$}
	\put(25,95){\small $\hat{x}$}
	\put(28,82){\small $\hat{z}$}
	
	\put(17,30){\small $\gamma_p$}
	\put(13,15){\small $e^{-}$}
	\put(40,15){\small $e^{+}$}	
	
	\put(7,50){\rotatebox{-10}{\small polarized $\gamma_p$}}	
	\put(70,39){\rotatebox{-10}{\small asymmetric laser}}	
	
	\put(57,136){{\footnotesize (a1)} }
	\put(105,134){\footnotesize (a2)}
	\put(126,93){\footnotesize (a3)}
	\put(85,130){\small $\xi_{3}$}	
	\put(65,95){\small $\xi_{1}$}
	\put(105,100){\small $\xi_{2}$}
	\put(60,76){\small Poincar\'{e} sphere}

	\put(143,123){\footnotesize (b1)}
	\put(204,120){\footnotesize (b2)}
	\put(227,86){\footnotesize (b3)}
	\put(178,122){\small $\hat{y}$}	
	\put(191,110){\small $\hat{x}$}
	\put(193,103){\small $\hat{z}$}
	\put(166,105){\small $\overline{{\bm S}}_+$}
	\put(150,67){\small Polarization states of $e^+$}
	
	\put(155,20){\rotatebox{-10}{\small Magnet}}
	
	\end{picture}	
	\caption{
		Scenario of generation of ASP lepton beams via nonlinear BW process. A LP asymmetric laser pulse (propagating along $-\hat{z}$ direction and polarizing along $\hat{x}$ axis)  head-on collides with polarized $\gamma$ photons ($\gamma_p$) to create ASP electron and positron beams. 	
		(a1)-(a3) indicate LP, CP and EP $\gamma$ photons, respectively, and (b1)-(b3) show average polarizations of created positrons $\overline{\bm S}_+$ corresponding to (a1)-(a3), respectively. The red-solid arrow and ellipsoid indicate the direction and amplitude of  $\overline{\bm S}_+$, respectively. The red-dashed arrows in (b1) ($\parallel\hat{{y}}$) and (b2) ($\parallel\hat{{z}}$) indicate particular cases of neglecting the polarization of $\gamma$ photons and employing symmetric laser field, respectively.
 	} \label{fig1}
\end{figure}
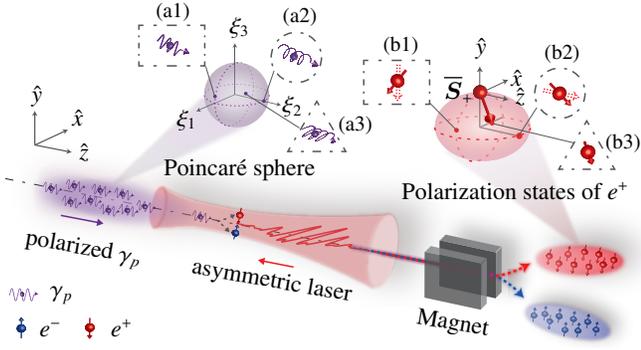

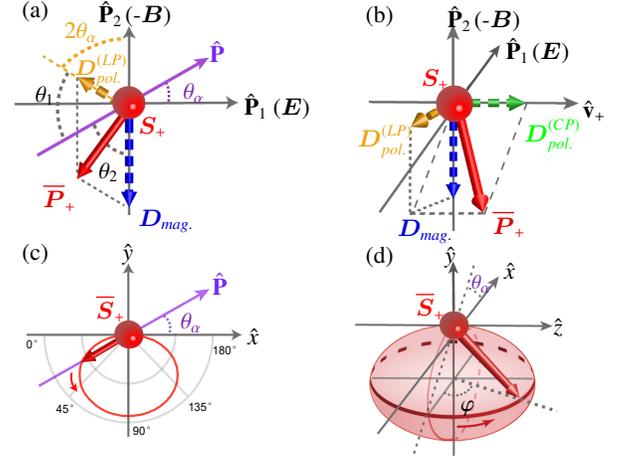
\begin{figure}[t]
	\centering
	\setlength{\abovecaptionskip}{-0.7cm}  	
	\includegraphics[width=1.0\linewidth]{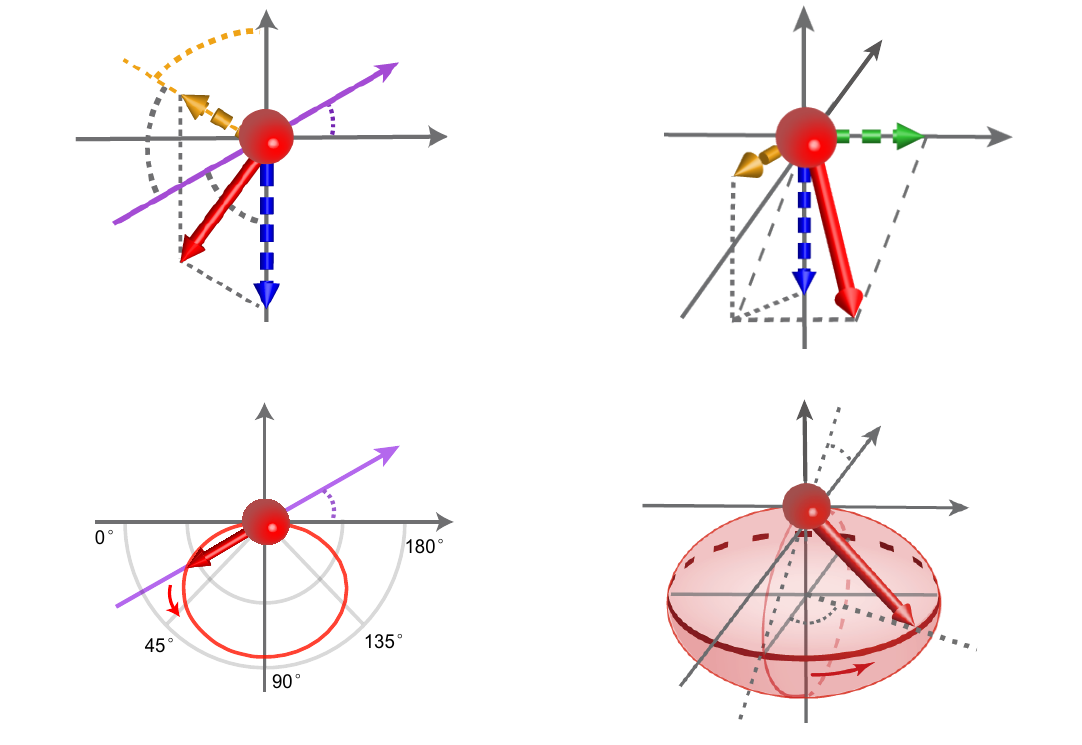}
	\begin{picture}(300,25)
		\put(20,197){\small (a)}
		\put(104,160){\small $\hat{\bf P}_1$}
		\put(115,160){\small (${\bm E}$)}
		\put(49,195){\small $\hat{\bf P}_2$}	
		\put(59,195){\small (-${\bm B}$)}
		\put(91,180){\small \textcolor[RGB]{116,37,248}{$\hat{\bf P}$}}
		\put(81,166){\small {\textcolor[RGB]{116,37,178}{$\theta_\alpha$} }}
		\put(25,165){\small $\theta_1$ }
		\put(49,136){\small $\theta_2$ }
		\put(64,153){\small {\color{red}{${\bm S}_+$} }}	
		\put(36,189){\small \textcolor[RGB]{239,163,23}{$2\theta_\alpha$}}
		\put(64,118){\small \textcolor[RGB]{0,0,225}{${\bm D}_{mag.}$}}
		\put(28,125){\small \textcolor[RGB]{225,0,0}{${\overline{\bm P}}_{+}$}}
		\put(40,175){\footnotesize  \textcolor[RGB]{239,163,23}{${\bm D}_{pol.}^{(LP)}$}}
		
		\put(20,105){\small (c)}
		\put(58,105){\small $\hat{y}$}
		\put(105,73){\small $\hat{x}$}
		\put(80,79){\small {\textcolor[RGB]{116,37,178}{$\theta_\alpha$} }}
		\put(48,83){\small {\color{red}{$\overline{\bm S}_+$} }}	
		\put(92,92){\small \textcolor[RGB]{116,37,248}{$\hat{\bf P}$}}

		\put(150,195){\small (b)}	
		\put(162,114){\small \textcolor[RGB]{0,0,225}{${\bm D}_{mag.}$}}
		\put(198,115){\small \textcolor[RGB]{225,0,0}{${\overline{\bm P}}_{+}$}}
		\put(147,148){\small \textcolor[RGB]{239,163,23}{${\bm D}_{pol.}^{(LP)}$}}
		\put(210,150){\small \textcolor[RGB]{0,255,0}{${\bm D}_{pol.}^{(CP)}$}}
		\put(171,171){\small {\color{red}{${\bm S}_+$} }}
		\put(203,181){\small $\hat{\bf P}_1$}
		\put(213,181){\small (${\bm E}$)}
		\put(181,193){\small $\hat{\bf P}_2$}	
		\put(191,193){\small (-${\bm B}$)}
		\put(232,159){\small $\hat{{\bf v}}_+$}
		\put(150,104){\small (d)}
		\put(180,104){\small $\hat{y}$}

		\put(202,98){\small $\hat{x}$}

		\put(221,77){\small $\hat{z}$}

		\put(169,84){\small {\color{red}{$\overline{\bm S}_+$} }}	
		\put(186,48){\small $\varphi$}
		\put(190,94){\footnotesize  {\textcolor[RGB]{116,37,178}{$\theta_\alpha$} }}	
	\end{picture}	
	
	\caption
	{ (a) and (b): Polarization of sample positron ${\bm S}_+$ created by LP and EP $\gamma$ photons, respectively. In our interaction scheme [see Fig.~\ref{fig1}] we employ $\hat{\bf P}_1 = \hat{\bm E} \approx \hat{{\bf a}}_+$ and $\hat{\bf P}_2 = -\hat{\bm B} \approx \hat{{\bm b}}_+$. ${\overline{\bm P}}_{+}$ indicates the direction of the instantaneous SQA with two transverse components ${\bm D}_{mag}$, ${\bm D}_{pol}^{(LP)}$ and one longitudinal component $D_{pol}^{(CP)}$ [see Eq.(\ref{sqa})]. For LP $\gamma$ photon in (a), $D_{pol}^{(CP)}=0$ and $\theta_\alpha$ indicates the polarization angle to $\hat{\bf P}_1$. $\theta_1$ and $\theta_2$ are the angles of $\hat{\bf P}$ to ${\bm D}_{pol.}^{(LP)}$ and ${\bm D}_{mag.}$, respectively.
		(c) and (d): Average polarization of positrons $\overline{\bm S}_+$ created by LP and EP $\gamma$ photons, respectively. The red arrow and circle (ellipsoid) indicate the direction and amplitude of $\overline{\bm S}_+$ [see Eq.~(\ref{ave_S})], respectively.
	} 
	\label{fig2}
\end{figure}

\begin{figure}[t]
	\centering
	\setlength{\abovecaptionskip}{-0.5cm}
	\includegraphics[width=1.0\linewidth]{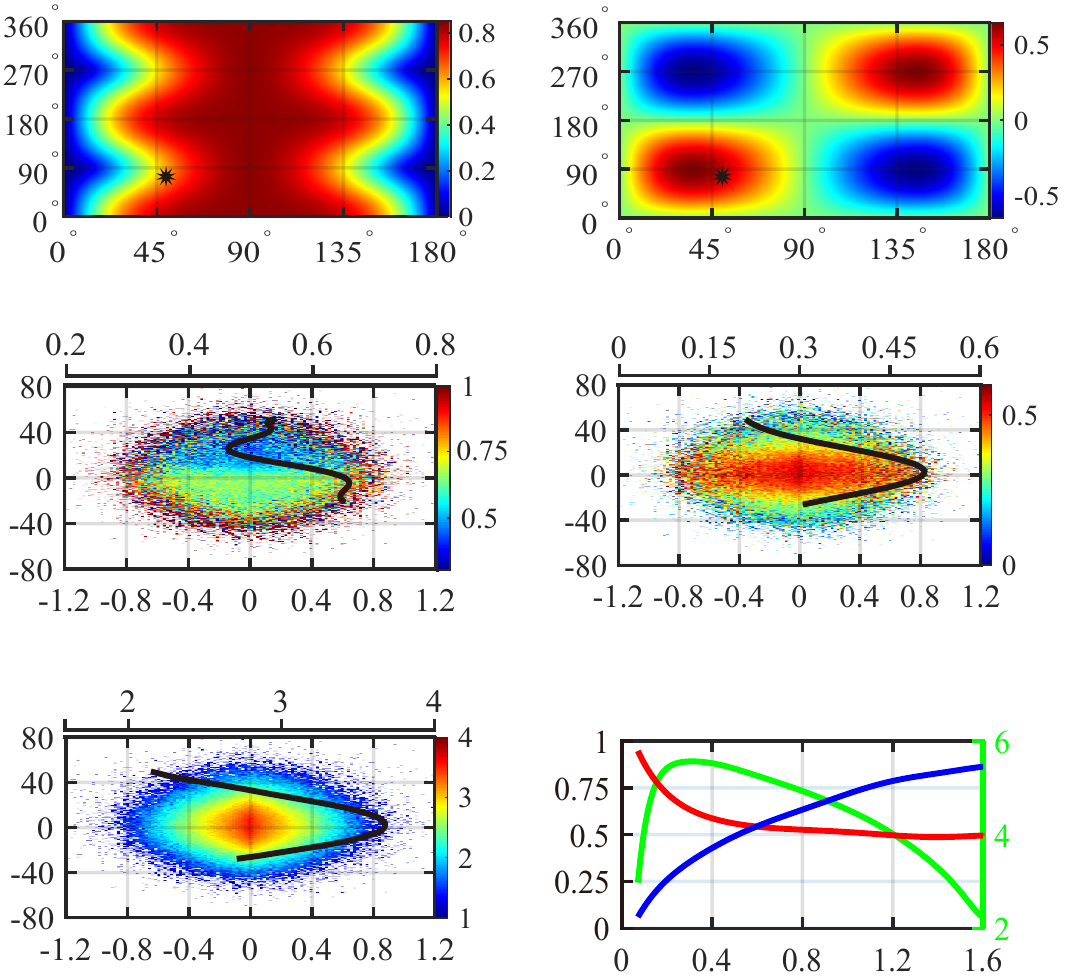}
	\begin{picture}(300,25)
		\put(18,153){\small (c)}
		\put(45,103.5){\small $\theta_y$ (mrad)}
		\put(-5,130){\rotatebox{90}{\small $\theta_x$ (mrad)} }
		\put(55,174){\small $\overline{\bm S}_T$}
		
		\put(213,153){\small (d)}	
		\put(174,103.5){\small $\theta_y$ (mrad)}
		\put(123,130){\rotatebox{90}{\small $\theta_x$ (mrad)}}
		\put(182,174){\small $\overline{\bm S}_L$}
		
		\put(18,72){\small (e)}
		\put(45,20){\small $\theta_y$ (mrad)} 
		\put(-5,49){\rotatebox{90}{\small $\theta_x$ (mrad)}}
		\put(30,94){\small log$_{10}$(d$N_+$/d$\theta_x$)}
		
		\put(215,65){\small (f)}	
		\put(175,18){\small $\varepsilon_{+}$(GeV)}
		\put(117,47){\rotatebox{90}{\small {\textcolor{red}{$\overline{\bm S}_T$}}{\color{black}$|$}{\textcolor{blue}{$\overline{\bm S}_L$}}}}
		\put(235,85){\rotatebox{-90}{\small {\color{green} log$_{10}$(d$N_+$/d$\varepsilon_{+}$)}}}
		
		\put(18,238){\small \textcolor{red}{(a)}}	
		\put(56,185){\small $\theta_\alpha$}
		\put(-7,218){\rotatebox{90}{\small $\theta_{\beta}$}}
		\put(42,208){{\scriptsize  {$63.3\%$}}}
		
		\put(215,238){\small (b)}	
		\put(183,186){\small $\theta_\alpha$}
		\put(118,218){\rotatebox{90}{\small $\theta_{\beta}$}}
		\put(170,208){{\scriptsize  {$53.5\%$}}}
		
	\end{picture}	
	\caption
	{	(a) and (b):  $\overline{\bm S}_T$ and $\overline{\bm S}_L$ of positrons with respect to $\theta_\alpha$ and $\theta_{\beta}$, respectively, analytically estimated by Eq.~(\ref{ave_S}) with ${\cal A}_{field} = 0.8378$.
		The black points correspond to ($\theta_\alpha=50^{\circ}$, $\theta_{\beta}=70^{\circ}$). (c)-(f) are the numerical simulation results with ($\theta_\alpha=50^{\circ}$, $\theta_{\beta}=70^{\circ}$).
		(c)-(e): $\overline{\bm S}_T$, $\overline{\bm S}_L$  and  the angle-resolved positron density ${\rm log}_{10}({\rm d}^2N_{+}/{\rm d}\theta_x{\rm d}\theta_y)$ (rad$^{-2}$) with respect to the deflection angles $\theta_x$ = arctan($p_{+,x}/p_{+,z}$) and 	$\theta_y$ = arctan($p_{+,y}/p_{+,z}$), respectively. The black curves indicate $\overline{\bm S}_T$, $\overline{\bm S}_L$ and ${\rm log}_{10}({\rm d}N_+/{\rm d}\theta_x)$ (mrad$^{-1}$) summing over $\theta_y$ vs $\theta_x$, respectively.
		(f) $\overline{\bm S}_T$ (red), $\overline{\bm S}_L$ (blue) and the energy-resolved positron density ${\rm log}_{10}({\rm d}N_+/{\rm d}\varepsilon_{+})$ (GeV$^{-1}$) (green)  vs $\varepsilon_{+}$.
		Other parameters are given in the text.
	} \label{fig3}
\end{figure}
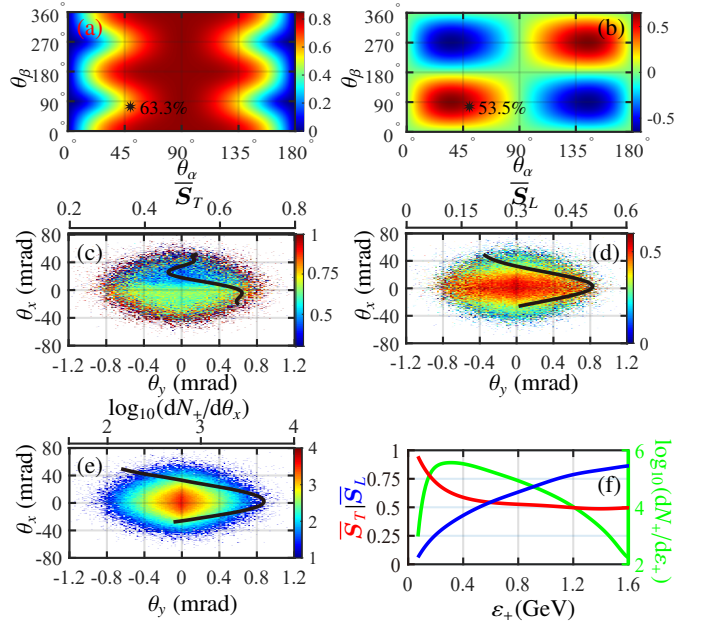

In this Letter, the generation of ASP GeV lepton beams has been  investigated in the interaction of polarized $\gamma$ photons with a counter-propogating ultraintense linearly polarized (LP) asymmetric laser pulse [see the interaction scenario in Fig.~\ref{fig1}]. We develop a fully spin-resolved semi-classical Monte Carlo algorithm to describe photon-polarization-dependent pair production and polarization in nonlinear BW process. 
We find that the polarization of created pairs is simultaneously determined by the polarization of parent $\gamma$ photons, the polarization and asymmetry of scattering laser field, due to the spin angular momentum transfer and the asymmetric spin-dependent pair production probabilities, respectively [see details in Fig.~\ref{fig2} and Eq.~(\ref{ave_S})]. As  employing unpolarized $\gamma$ photons or  ignoring the polarization of parent $\gamma$ photons  the pair polarization will rely on the laser polarization and asymmetry [see Fig.~\ref{fig1}(b1)], and as employing a symmetric laser field the transverse polarization of the total beam will be suppressed (since the polarization directions in adjacent half laser cycles are opposite) and only the longitudinal polarization can be retained [see Fig.~\ref{fig1}(b2)].
Our simulations show that dense GeV  lepton beams with adjustable polarization degree up to about 80\% can be obtained with currently achievable laser facilities to the benefit of many unique applications [see details in Fig.~\ref{fig3}].

In strong laser field, a rich pair yield via nonlinear BW process requires the nonlinear QED parameter $\chi_{\gamma}\equiv |e|\sqrt{-(F_{\mu\nu}k_{\gamma}^{\nu})^2}/m^3\gtrsim1$  \cite{Ritus_1985, Baier1998}, and the created pairs may further emit photons via nonlinear Compton scattering, which is unnegligible as another nonlinear QED parameter $\chi_e\equiv |e|\sqrt{-(F_{\mu\nu}p^{\nu})^2}/m^3\gtrsim 1$ \cite{Ritus_1985}. Here, $e$ and $m$ are the electron charge and mass, respectively, $k_\gamma$ and $p$ the 4-momenta of $\gamma$ photon and positron (electron), respectively, and $F_{\mu\nu}$ the field tensor. Relativistic units with $c=\hbar=1$ are used throughout. 
The photon polarization can be characterized by the unit vector $\hat{\bf P} = {\rm cos}(\theta_\alpha)\hat{\bf P}_1+{\rm sin}(\theta_\alpha)\hat{\bf P}_2\cdot e^{i\theta_\beta}$, and corresponding Stokes parameters are $(\xi_{1}, \xi_{2}, \xi_{3}) = [{\rm sin}(2\theta_\alpha){\rm cos}(\theta_\beta)$, ${\rm sin}(2\theta_\alpha){\rm sin}(\theta_\beta)$, ${\rm cos}(2\theta_\alpha)]$ \cite{McMaster_1961,Boyarkin2011}.
Here $\hat{{\bf P}}_1$ and $\hat{{\bf P}}_2$ are two orthogonal basis vectors, $\theta_\alpha$  the polarization angle, $\theta_\beta$ the absolute phase, $\xi_{1}$ and $\xi_{3}$ describe the linear polarizations,
and $\xi_{2}$   circular polarization.  
The fully spin-resolved pair production probability $W_{pair}$ has been calculated 
via the QED operator method of Baier-Katkov-Fadin \cite{Baier_1973} in the local constant field approximation  \cite{Ritus_1985,Baier1998,Piazza_2018,Ilderton2019prd, piazza2019, Seipt_2020} (valid at the invariant field parameter $a_0 = |e|E_0/m\omega \gg1$, where $E_0$ and $\omega_0$ are the laser field amplitude  and frequency, respectively); see the complex expression of $W_{pair}$ in \cite{supplemental}.

Let's first summarize our methods of numerical simulation and analytical estimation. Note that in nonlinear BW process the polarization of electrons is similar with that of positrons, thus for simplicity  we only discuss the case of positrons below. 
To study the  positron polarization ${\bm S}_+$, we first sum over the electron polarization ${\bm S}_-$  in $W_{pair}$, and the probability  relying on ${\bm S}_+$ is simplified as:
\begin{eqnarray}
	\label{BW_pos}
	\frac{{\rm d}^2W_{pair}^{+}}{{\rm d}\varepsilon_{+}{\rm d}t} = W_{0}(C+{\bm S}_{+}\cdot{\bm D}),
\end{eqnarray}
where $W_{0} = \alpha m^2/\left(\sqrt{3}\pi\varepsilon_{\gamma}^2\right)$, $C={\rm IntK}_{\frac{1}{3}}(\rho) + \frac{\varepsilon_{-}^2+\varepsilon_{+}^2}{\varepsilon_{-}\varepsilon_{+}}{\rm K}_{\frac{2}{3}}(\rho) - \xi_{3}{\rm K}_{\frac{2}{3}}(\rho)$, 
	${\bm D} =\left(\xi_{3}\frac{\varepsilon_{\gamma}}{\varepsilon_{-}}-\frac{\varepsilon_{\gamma}}{\varepsilon_{+}}\right){\rm K}_{\frac{1}{3}}(\rho)\hat{{\bm b}}_+ - \xi_{1}\frac{\varepsilon_{\gamma}}{\varepsilon_{-}}{\rm K}_{\frac{1}{3}}(\rho)\hat{{\bf a}}_+ + \xi_{2}\bigg[ \frac{\varepsilon_{+}^2-\varepsilon_{-}^2}{\varepsilon_{-}\varepsilon_{+}}{\rm K}_{\frac{2}{3}}(\rho)+\frac{\varepsilon_{\gamma}}{\varepsilon_{+}}{\rm IntK}_{\frac{1}{3}}(\rho)\bigg]\hat{\bf v}_+$,   $\hat{\bm b}_+=\hat{\bf v}_+\times\hat{{\bf a}}_+$ $\approx-\hat{\bm k}_\gamma\times \hat{\bm E}= -\hat{\bm B}$ is an unit vector anti-parallel to the magnetic field ${\bm B}$ (with direction vector $\hat{\bm B}$) in the rest frame of positron, ${\bm E}$ the electric field with direction vector $\hat{\bm E}$, $\hat{\bf v}_+$ and $\hat{\bf a}_+$ the unit vectors of the positron velocity and acceleration, respectively, $\rho = 2\varepsilon_{\gamma}^2/\left(3\chi_{\gamma}\varepsilon_{-}\varepsilon_{+}\right)$, ${\rm IntK}_{\frac{1}{3}}(\rho)\equiv \int_{\rho}^{\infty} {\rm d}z {\rm K}_{\frac{1}{3}}(z)$, ${\rm K}_n$  the $n$-order modified Bessel function of the second kind, $\alpha$ the fine structure constant, $\varepsilon_{\gamma}$, $\varepsilon_{-}$ and $\varepsilon_{+}$ the energies of parent $\gamma$ photon, created electron and positron, respectively, with $\varepsilon_{\gamma}= \varepsilon_{+}+\varepsilon_{-}$. When a $\gamma$ photon decays into a pair, the positron spin state is  collapsed into one of its basis  states defined by the instantaneous spin quantization axis (SQA) along the energy-resolved average polarization vector~\cite{Baier_1973}: 
	\begin{eqnarray}
	{\rm SQA}\parallel{\overline{\bm P}_{+}}={\bm D}/C, \label{SQA}
	\end{eqnarray}
which can be rewritten as
\begin{eqnarray}\label{sqa}
	{\overline{\bm P}}_{+} &=& \left[ {\bm D}_{mag.}+{\bm D}_{pol.}^{(LP)}+{\bm D}_{pol.}^{(CP)}\right]/C\nonumber\\
	& =& \left[|{\bm D}_{mag.}|\hat{\bm D}_{mag.}+|{\bm D}_{pol.}^{(LP)}|\hat{\bm D}_{pol.}^{(LP)}+|{\bm D}_{pol.}^{(CP)}|\hat{\bm D}_{pol.}^{(CP)}\right]/C,
\end{eqnarray}
where $\hat{\bm D}_{mag.} = -\hat{{\bf b}}_+ \approx \hat{\bm B}$,
 $\hat{\bm D}_{pol.}^{(LP)} = \xi_{3}\hat{{\bm b}}_+-\xi_{1}\hat{{\bf a}}_+$ and $\hat{\bm D}_{pol.}^{(CP)} = \xi_{2} \hat{{\bf v}}_+$ rely on the magnetic field $\hat{\bm B}$, linear polarizations $\xi_1$ and $\xi_3$, and circular polarization $\xi_2$, respectively, with corresponding factors
 $|{\bm D}_{mag.}| = \frac{\varepsilon_{\gamma}}{\varepsilon_{+}}{\rm K}_{\frac{1}{3}}(\rho)$, 
 $|{\bm D}_{pol.}^{(LP)}| = \frac{\varepsilon_{\gamma}}{\varepsilon_{-}}{\rm K}_{\frac{1}{3}}(\rho)$ and $|{\bm D}_{pol.}^{(CP)}| = \frac{\varepsilon_{+}^2-\varepsilon_{-}^2}{\varepsilon_{-}\varepsilon_{+}}{\rm K}_{\frac{2}{3}}(\rho)+\frac{\varepsilon_{\gamma}}{\varepsilon_{+}}{\rm IntK}_{\frac{1}{3}}(\rho)$, respectively. 
 As the photon polarization is ignored, ${\rm SQA}\parallel \hat{\bm D}_{mag.}\parallel \hat{\bm B}$, i.e., the positrons are polarized along the magnetic field ${\bm B}$~\cite{Chen_2019} [see Fig.~\ref{fig1}(b1)].  The polarization geometries of positrons created by LP and EP $\gamma$ photons are illustrated in Figs.~\ref{fig2}(a) and (b), respectively. For LP $\gamma$ photon, $\theta_\beta=0$, $\hat{\bm D}_{pol.}^{(CP)}=0$, $(\xi_{1}, \xi_{2}, \xi_{3}) = [{\rm sin}(2\theta_\alpha)$, 0, ${\rm cos}(2\theta_\alpha)]$, and $\hat{\bm D}_{pol.}^{(LP)} = \xi_{3}\hat{{\bf b}}_+-\xi_{1}\hat{{\bf a}}_+ = {\rm cos}(2\theta_\alpha)\hat{\bf P}_2 - {\rm sin}(2\theta_\alpha)\hat{\bf P}_1$. 
 For general EP $\gamma$ photon with $\xi_2\neq 0$, the longitudinal polarization component must be taken into account.
 
\begin{figure}[t]
	\centering
	\setlength{\abovecaptionskip}{-0.7cm}  	
	\includegraphics[width=1.0\linewidth]{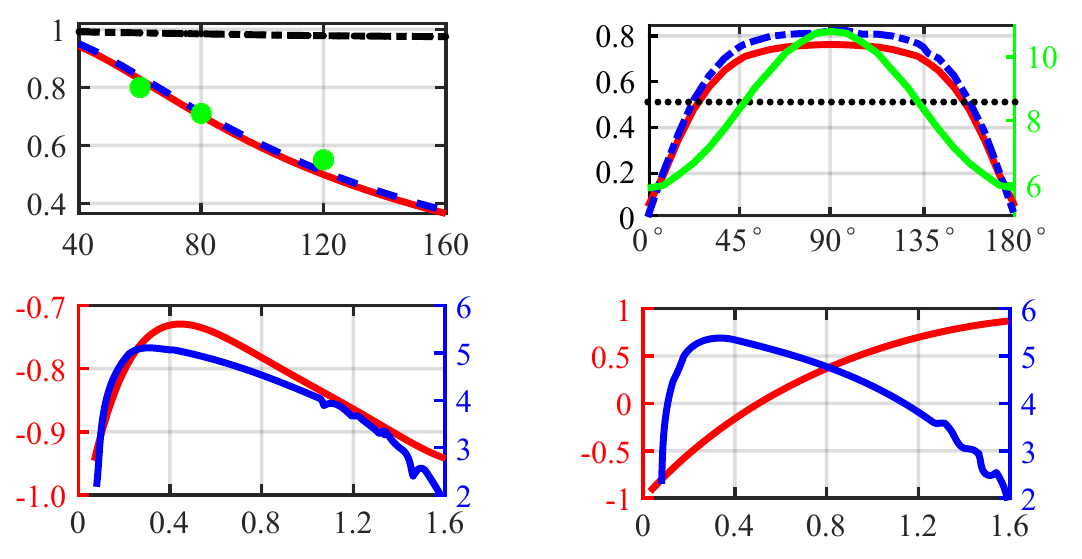}
	\begin{picture}(300,25)
	
		\put(150,138){\small (b)}
		\put(186,87){\small $\theta_\alpha$}
		\put(124,123){\rotatebox{90}{\small $\overline{\bm S}_T$}}
		\put(242,139){\rotatebox{-90}{\small \textcolor{green}{ $N_+/N_\gamma$}}}
		\put(236,147){\small \textcolor{green}{$\times 10^{-2}$}}
		\put(178,124){\color{red}{\rule{0.037\linewidth}{1.4pt}}}  
		
		\put(178,118){\color{blue}{\rule{0.011\linewidth}{1.4pt}}}  
		\put(182,118){\color{blue}{\rule{0.005\linewidth}{1.4pt}}}    
		\put(184.5,118){\color{blue}{\rule{0.011\linewidth}{1.4pt}}} 
		
		\put(1748,111){\color{black}{\rule{0.005\linewidth}{1.4pt}}}  
		\put(180.5,111){\color{black}{\rule{0.005\linewidth}{1.4pt}}}  
		\put(183,111){\color{black}{\rule{0.005\linewidth}{1.4pt}}}  
		\put(185.5,111){\color{black}{\rule{0.005\linewidth}{1.4pt}}}  
		
		\put(188,122){\color{red}{\scriptsize MC}} 
		\put(188,116){\color{blue}{\scriptsize No Rad.}} 
		\put(188.5,109){\color{black}{\scriptsize Exc. Pol.}}

		\put(20,128){\small (a)}
		\put(56,90){\small $a_1$}
		\put(73,138){\color{red}{\rule{0.04\linewidth}{1.4pt}}}  
		\put(83,136){\color{red}{\scriptsize $\overline{\bm S}_{T}$}} 
		
		\put(73,131){\color{blue}{\rule{0.015\linewidth}{1.4pt}}}    
		\put(78,131){\color{blue}{\rule{0.015\linewidth}{1.4pt}}}  
		\put(83,129){\color{blue}{\scriptsize ${\cal A}_{field}$}} 
		
		\put(73,124){\color{black}{\rule{0.012\linewidth}{1.4pt}}}  
		\put(77,124){\color{black}{\rule{0.005\linewidth}{1.4pt}}}  
		\put(79.5,124){\color{black}{\rule{0.012\linewidth}{1.4pt}}}  
		\put(83,122){\color{black}{\scriptsize ${\cal A}_{pol.}$}} 
		
		\put(33,117){\color{green}{\scriptsize $\overline{\bm S}_{T}$}} 
				
				\put(20,75){\small (c)} 
				\put(50,21){\small$\varepsilon_{+}$ (GeV) }
				\put(-6,60){\rotatebox{90}{\normalsize \textcolor{red}{\small ${\bm S}_T$ }}}
				\put(113,90){\rotatebox{-90}{\normalsize \textcolor{blue}{\small log$_{10}($d$N_+/$d$\varepsilon_{+})$ }}}
				\put(46,45){\small $\theta_{\alpha}=90^{\circ}$}
				
				\put(148,75){\small (d)} 
				\put(180,21){\small$\varepsilon_{+}$ (GeV) }
				\put(125,60){\rotatebox{90}{\normalsize \textcolor{red}{\small ${\bm S}_T$ }}}
				\put(240,90){\rotatebox{-90}{\normalsize \textcolor{blue}{\small log$_{10}($d$N_+/$d$\varepsilon_{+})$ }}}
				\put(175,45){\small $\theta_{\alpha}=0^{\circ}$}
	\end{picture}	
	
	\caption
	{   
		(a)  ${\cal A}_{pol.}$  (black-dash-dotted), ${\cal A}_{field}$ (blue-dashed) and  $\overline{\bm S}_{T}$ (red-solid) analytically calculated by Eq.~(\ref{ave_S}) vs  $a_1$ ($a_2=a_1/4$), and the green points $\overline{\bm S}_{T}$ are our numerical results. $\theta_\alpha=90^{\circ}$. At $a_1$ = 60, 80 and 120, respectively, the corresponding $\overline{\chi}_\gamma \approx$ 0.96, 1.20 and 1.63, respectively. 	
		(b) $\overline{\bm S}_T$ and the yield ratio of positrons $N_+/N_\gamma$ (green-solid) vs
$\theta_\alpha$. 
The red-solid, blue-dash-dotted and black-dotted  curves indicate the results of $\overline{\bm S}_T$ calculated by  our Monte Carlo method, artificially ignoring the radiative depolarization effects of the positrons propagating through the laser field, and excluding the polarization effects of parent $\gamma$ photons in nonlinear BW process, respectively. (c) and (d): ${\bm S}_T$ (red-solid) and log$_{10}($d$N_+/$d$\varepsilon_{+})$(GeV$^{-1}$) (blue)  vs $\varepsilon_{+}$ at $\theta_{\alpha}=90^{\circ}$ and $\theta_{\alpha}=0^{\circ}$, respectively. Here $\theta_\beta=0^{\circ}$, and other parameters are the same as those in Fig.~\ref{fig3}.		
	} 
	\label{fig4}
\end{figure}
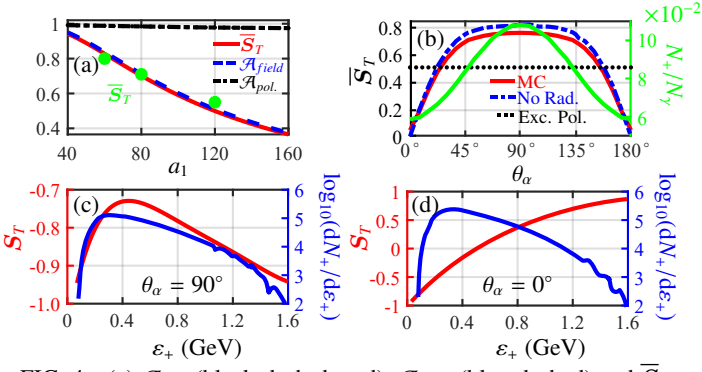

 For a positron beam, the average transverse polarizations $\overline{\bm S}_{T}$ in adjacent half laser cycles  are opposite and cancel each other out due to the laser field oscillation, and consequently, $\overline{\bm S}_{T}$ is proportional to the asymmetry of the laser field, which can be characterized by the relative deviation between the pair production probabilities in positive and negative  half cycles
 ${\cal A}_{field} = (W_{pair}^{+,pos.}-W_{pair}^{+,neg.})/(W_{pair}^{+,pos.}+W_{pair}^{+,neg.})$. Thus, one can estimate
\begin{eqnarray}
	\label{ave_S}
	\overline{{\bm S}}_{+}=\left\{{ {\cal A}_{field}\cdot \left[\overline{\bm D}_{mag.}+\overline{\bm D}_{pol.}^{(LP)}\right]+\overline{\bm D}_{pol.}^{(CP)}}\right\}/{\overline{C}},
\end{eqnarray}
where $\overline{\bm D}_{mag.} = \int_{0}^{\varepsilon_{\gamma}}{\bm D}_{mag.}{\rm d}\varepsilon_{+}$, $\overline{\bm D}_{pol.}^{(LP)} = \int_{0}^{\varepsilon_{\gamma}}{\bm D}_{pol.}^{(LP)}{\rm d}\varepsilon_{+}$, $\overline{\bm D}_{pol.}^{(CP)} = \int_{0}^{\varepsilon_{\gamma}}{\bm D}_{pol.}^{(CP)}{\rm d}\varepsilon_{+}$ and $\overline{ C} =\int_{0}^{\varepsilon_{\gamma}}C{\rm d}\varepsilon_{+}$. 
For LP $\gamma$ photons, $|\overline{\bm D}_{mag.}| = |\overline{\bm D}_{pol.}^{(LP)}|$ with ${\theta}_1={\theta}_2$ results in $\overline{\bm S}_+ \parallel -\hat{\bf P}$ [see Fig.~\ref{fig2}(c)];
for more general EP ones, as $\theta_\beta$ increases, $\overline{\bm S}_+$ will rotate anti-clockwise by an azimuth angle $\varphi$, which can be calculated by Eq.~(\ref{ave_S}) [see Fig.~\ref{fig2}(d)]. $\overline{\bm S}_{T} = {\cal A}_{field} {{\cal A}}_{pol.}$ is dominated by ${\cal A}_{field}$ 
with ${\cal A}_{pol.} = \left[\overline{\bm D}_{mag.}+\overline{\bm D}_{pol.}^{(LP)}\right]/\overline{C}$ [see Fig.~\ref{fig4}(a)], while the average longitudinal polarization $\overline{{\bm S}}_{L}$ is solely determined by $\overline{\bm D}_{pol.}^{(CP)}/\overline{C}\propto \xi_2$ as expected. 
In symmetric laser field with ${\cal A}_{field} = 0$, $\overline{\bm S}_{T}$ is very little and only $\overline{{\bm S}}_{L}$ can be obtained by employing longitudinally polarized $\gamma$ photons ($\xi_2 \neq 0$)~\cite{Li_2020_2}
[see Fig.~\ref{fig1}(b2)]. 
The momentum and spin dynamics of the pairs propagating through the laser field are calculated following a Monte Carlo algorithm~\cite{Wan_2020}, including the radiative depolarization effects and spin procession~\cite{Thomas_1926,Thomas_1927,Bargmann_1959}. See more details of our simulation method in \cite{supplemental}.
 
Then, we illustrate typical results of created ASP positron beams in Fig~\ref{fig3}. The  employed laser and $\gamma$ photon parameters  are as follows. A tightly focused LP bichromatic Gaussian laser pulse \cite{Salamin2002,supplemental} (a frequency-chirped laser pulse \cite{Galow_2011}  can work similarly) propagates along $-\hat{z}$ direction and polarizes along $\hat{x}$ axis [see Fig.~\ref{fig1}], with a phase difference $\Delta \phi = \pi/2$ to obtain the maximal field asymmetry, peak intensity  $I_0 \approx1.11\times10^{22}$ W/cm$^2$ (corresponding invariant field parameters $a_1 = 60$ and $a_2=15$),  wavelengths $\lambda_1=2\lambda_2 = 1\mu$m, pulse durations $\tau_1 = \tau_2 = 15T_1$ with periods $T_1=2T_2$, and focal radii $w_1 = w_2 =5\mu$m.
 This kind of laser pulse is currently feasible in petawatt laser facilities~\cite{Yoon2019,Danson_2019,Gales_2018}. 
 While,
a cylindrical polarized $\gamma$ photon beam propagates along $\hat{z}$ direction, with an initial energy $\varepsilon_\gamma=1.8$GeV, energy spread $\Delta \varepsilon_\gamma/\varepsilon_\gamma =0.06$,  angular divergence $ \Delta\theta_\gamma = 0.3$mrad, beam radius  $w_\gamma= 1\mu$m, beam length $L_\gamma = 5\mu$m, photon number $N_\gamma=10^6$ and density $n_\gamma\approx 6.37\times10^{16}$cm$^{-3}$ having a transversely Gaussian and longitudinally uniform distribution. Such a $\gamma$ photon beam may be generated via synchrotron radiation \cite{Alexander_2008,Baynham_2011}, bremsstrahlung \cite{Abbott_2016}, linear  \cite{Omori_2006} or nonlinear Compton scattering \cite{King_2020,Tang_2020, Ligammaray_2019}. The pair production is remarkable at these parameters since $\overline{\chi}_{\gamma}\approx 0.96$.

According to  Eq.~(\ref{ave_S}), the polarizations of created positron beam ($\overline{\bm S}_T$ and $\overline{\bm S}_L$) can be controlled by adjusting the polarization of parent $\gamma$ photons ($\theta_\alpha$ and $\theta_\beta$) [see Figs.~\ref{fig3}(a) and (b)], which indicates the spin angular momentum transfer from parent $\gamma$ photons to created pairs. $\overline{\bm S}_T \propto \sqrt{\xi_1^2+\xi_3^2} = \sqrt{{\rm sin}^2(2\theta_\alpha){\rm cos}^2(\theta_\beta)+{\rm cos}^2(2\theta_\alpha)}$  is mainly determined by $\theta_\alpha$ and can reach above 80\%, while
$\overline{\bm S}_L\propto \xi_2={\rm sin}(2\theta_\alpha){\rm sin}(\theta_\beta)$ periodically varies with respect to $\theta_\alpha$ and $\theta_\beta$ and its amplitude can achieve about 60\%. For a specific case with $\theta_\alpha = 50^{\circ}$ and $\theta_{\beta} = 70^{\circ}$, the analytical estimations are $\overline{\bm S}_T\approx 63.3\%$ and $\overline{\bm S}_L\approx 53.5\%$  [see the black-star points in Figs.~\ref{fig3}(a) and (b)], and corresponding numerical results are $\overline{\bm S}_T\approx 62.1\%$ and $\overline{\bm S}_L\approx 50.3\%$ \cite{supplemental}. The little deviations are derived from that in analytical estimations we employ a constant ${\cal A}_{field}$, which actually has 
 spatial and temporal profiles in our numerical simulations. As the created positrons propagate through the laser field, the average polarizations slightly decrease to  
$\overline{\bm S}_T\approx 59.4\%$ and $\overline{\bm S}_L\approx 44.8\%$ [see Figs.~\ref{fig3}(c) and (d)] due to the quantum radiative depolarization effects \cite{li2019prl} [see \cite{supplemental} and Fig.~\ref{fig4}(b)]. $\overline{\bm S}_T \propto {\cal A}_{field}$ is asymmetric in angular distribution due to asymmetric ${\cal A}_{field}$, which 
doesn't affect the symmetry of $\overline{\bm S}_L\propto\xi_2$.
The yield rate of positrons $N_+/N_\gamma \approx 0.1$ is much higher than that of the common method employing Bethe-Heitler pair producation ($\sim 10^{-3}-10^{-4}$) \cite{Omori_2006,Alexander_2008,abbott2016prl}, since $N_+\sim \alpha a_0$ is rather large in ultraintense laser field
\cite{Ritus_1985,Baier_1973}; see Fig.~\ref{fig3}(e).
The flux of the positron beam is approximately $3.0\times 10^{19}$/s  with duration $\tau_+ \approx \tau_\gamma$ (due to the relativistic effect).
The emittance $\epsilon \approx w_+\theta_{div.}\sim 10^{-2}$ mm$\cdot$mrad  fulfills the experimental requirements of the beam injectors \cite{Artru_2008}, with  radius $w_+ \approx w_\gamma=1\mu$m and angular divergence $\theta_{div.} \sim 30$mrad [FWHM in Fig.~\ref{fig3}(e)].
Due to the stochastic effects of the pair production and further radiation, the energy of the positron beam spreads with a density peak at $\varepsilon_{+}\approx0.3$GeV [see Fig.~\ref{fig3}(f)]. 
With the increase of $\varepsilon_{+}$, 
$\overline{\bm S}_T$ ($\overline{\bm S}_L$) monotonically decreases (increases) from above 90\% (0\%) to  about $50\%$ (above $85\%$), since the pair polarization is mainly determined by the polarization of the laser (parent $\gamma$ photons) at low (high) $\varepsilon_{+}$ (see \cite{supplemental}). 
For $\varepsilon_{+}=$ 0.4GeV, 0.8GeV and 1.2GeV, respectively, corresponding $\overline{\bm S}_T$ ($\overline{\bm S}_L$) is about 58.7\%, 52.6\% and 49.8\% (42.6\%, 63.4\% and 78.8\%), respectively, and brilliance about 4.6$\times$10$^{20}$, 1.4$\times$10$^{20}$ and 0.5$\times$10$^{20}$ positrons/(s mm$^{2}$ mrad$^{2}$ 0.1\% BW), respectively, with angular divergence (FWHM) of about 24.9 mrad$^2$, 15.9 mrad$^2$ and 13.0 mrad$^2$, respectively. 

We underline that as the ultra-relativistic laser intensity $a_0 (\sim a_1) \propto \chi_\gamma$ continuously increases,  $W_{pair}^{+,pos.}$ and $W_{pair}^{+,neg.}$ are both gradually approaching the saturation threshold, and thus $\overline{\bm S}_T \propto {\cal A}_{field}$ decreases continuously [see Fig.~\ref{fig4}(a)]. As employing unpolarized $\gamma$ photons or artificially ignoring the polarization effects of $\gamma$ photons in nonlinear BW process, for given parameters $\overline{\bm S}_T$ is only about 50.8\%, however, employing polarized $\gamma$ photons $\overline{\bm S}_T$ can achieve up to 76.2\% with a peak yield rate $N_+/N_\gamma\approx0.11$ at $\theta_\alpha=90^\circ$. And in a broad range of $45^{\circ} \lesssim \theta_\alpha \lesssim 135^{\circ}$ one can obtain $\overline{\bm S}_T \gtrsim 60\%$ with $N_+/N_\gamma\gtrsim 0.08$ [see Fig.~\ref{fig4}(b)]. The energy-resolved $\overline{\bm S}_T$ and densities at $\theta_\alpha=90^\circ$ and $0^\circ$ are represented in Figs.~\ref{fig4}(c) and (d), respectively. It's interesting that at $\theta_\alpha=0^\circ$ even though  $\overline{\bm S}_T$ is very little, highly polarized positron beams can still be obtained by energy choosing by magnets.

For experimental feasibility, the impact of the laser and $\gamma$ photon parameters (e.g., the laser intensity, pulse duration, colliding angle, and the angular spreading, energy and energy spreading  of the $\gamma$ photon beam) on the positron polarization  is investigated, and the results keep uniform (see~\cite{supplemental}). Moreover, for a more general case: an electron beam head-on collides with a LP bichromatic laser pulse to generate positrons, the positrons are only weakly transversely polarized, since the polarization of intermediate $\gamma$ photons is nearly parallel to that of the laser field
\cite{supplemental}.

In conclusion, we reveal the fully spin-resolved pair polarization mechanism in nonlinear BW process, which   can be observed by the average polarization vector  in an asymmetric (e.g. well known bichromatic and frequency-chirped) laser field. And we  put forward an efficient method to generate dense ASP GeV  lepton beams with the  polarization degree up to about $80\%$ with currently achieveable petawatt laser facilities \cite{Yoon2019,Danson_2019,Gales_2018}, 
which have unique applications for high-energy physics and particle physics, in particular, 
as injectors of the polarized $e^{+}e^{-}$ colliders for searching for BSM new physics
 \cite{Moortgat2008,Ananthanarayan_2004, Ananthanarayan_2005,Ananthanarayan2018,
Herczeg2003,Dass1977,Burgess1991,
Ananthanarayan2004,Harlander1997,Godbole_2006,
Rizzo2003,Choi2015}. 
\\

{\textbf{Acknowledgement:}} This work is supported by the National Natural Science Foundation of China (Grants Nos. 12022506, 11874295, 11875219 and 11905169), and the National Key R\&D Program of China (Grant No. 2018YFA0404801).


\bibliography{QEDspin}

\begin{thebibliography}{93}%
\makeatletter
\providecommand \@ifxundefined [1]{%
 \@ifx{#1\undefined}
}%
\providecommand \@ifnum [1]{%
 \ifnum #1\expandafter \@firstoftwo
 \else \expandafter \@secondoftwo
 \fi
}%
\providecommand \@ifx [1]{%
 \ifx #1\expandafter \@firstoftwo
 \else \expandafter \@secondoftwo
 \fi
}%
\providecommand \natexlab [1]{#1}%
\providecommand \enquote  [1]{``#1''}%
\providecommand \bibnamefont  [1]{#1}%
\providecommand \bibfnamefont [1]{#1}%
\providecommand \citenamefont [1]{#1}%
\providecommand \href@noop [0]{\@secondoftwo}%
\providecommand \href [0]{\begingroup \@sanitize@url \@href}%
\providecommand \@href[1]{\@@startlink{#1}\@@href}%
\providecommand \@@href[1]{\endgroup#1\@@endlink}%
\providecommand \@sanitize@url [0]{\catcode `\\12\catcode `\$12\catcode
  `\&12\catcode `\#12\catcode `\^12\catcode `\_12\catcode `\%12\relax}%
\providecommand \@@startlink[1]{}%
\providecommand \@@endlink[0]{}%
\providecommand \url  [0]{\begingroup\@sanitize@url \@url }%
\providecommand \@url [1]{\endgroup\@href {#1}{\urlprefix }}%
\providecommand \urlprefix  [0]{URL }%
\providecommand \Eprint [0]{\href }%
\providecommand \doibase [0]{http://dx.doi.org/}%
\providecommand \selectlanguage [0]{\@gobble}%
\providecommand \bibinfo  [0]{\@secondoftwo}%
\providecommand \bibfield  [0]{\@secondoftwo}%
\providecommand \translation [1]{[#1]}%
\providecommand \BibitemOpen [0]{}%
\providecommand \bibitemStop [0]{}%
\providecommand \bibitemNoStop [0]{.\EOS\space}%
\providecommand \EOS [0]{\spacefactor3000\relax}%
\providecommand \BibitemShut  [1]{\csname bibitem#1\endcsname}%
\let\auto@bib@innerbib\@empty
\bibitem [{\citenamefont {Wardle}\ \emph {et~al.}(1998)\citenamefont {Wardle},
  \citenamefont {Homan}, \citenamefont {Ojha},\ and\ \citenamefont
  {Roberts}}]{Wardle_1998}%
  \BibitemOpen
  \bibfield  {author} {\bibinfo {author} {\bibfnamefont {J.~F.~C.}\
  \bibnamefont {Wardle}}, \bibinfo {author} {\bibfnamefont {D.~C.}\
  \bibnamefont {Homan}}, \bibinfo {author} {\bibfnamefont {R.}~\bibnamefont
  {Ojha}}, \ and\ \bibinfo {author} {\bibfnamefont {D.~H.}\ \bibnamefont
  {Roberts}},\ }\bibfield  {title} {\enquote {\bibinfo {title}
  {Electron-positron jets associated with the quasar 3c279},}\ }\href@noop {}
  {\bibfield  {journal} {\bibinfo  {journal} {Nature}\ ,\ \bibinfo {pages}
  {457--461}} (\bibinfo {year} {1998})}\BibitemShut {NoStop}%
\bibitem [{\citenamefont {\ifmmode \check{Z}\else
  \v{Z}\fi{}uti\ifmmode~\acute{c}\else \'{c}\fi{}}\ \emph
  {et~al.}(2004)\citenamefont {\ifmmode \check{Z}\else
  \v{Z}\fi{}uti\ifmmode~\acute{c}\else \'{c}\fi{}}, \citenamefont {Fabian},\
  and\ \citenamefont {Das~Sarma}}]{RevModPhys.76.323}%
  \BibitemOpen
  \bibfield  {author} {\bibinfo {author} {\bibfnamefont {Igor}\ \bibnamefont
  {\ifmmode \check{Z}\else \v{Z}\fi{}uti\ifmmode~\acute{c}\else \'{c}\fi{}}},
  \bibinfo {author} {\bibfnamefont {Jaroslav}\ \bibnamefont {Fabian}}, \ and\
  \bibinfo {author} {\bibfnamefont {S.}~\bibnamefont {Das~Sarma}},\ }\bibfield
  {title} {\enquote {\bibinfo {title} {Spintronics: Fundamentals and
  applications},}\ }\href {\doibase 10.1103/RevModPhys.76.323} {\bibfield
  {journal} {\bibinfo  {journal} {Rev. Mod. Phys.}\ }\textbf {\bibinfo {volume}
  {76}},\ \bibinfo {pages} {323--410} (\bibinfo {year} {2004})}\BibitemShut
  {NoStop}%
\bibitem [{\citenamefont {The BESIII~Collaboration.}(2019)}]{Ablikim2018zay}%
  \BibitemOpen
  \bibfield  {author} {\bibinfo {author} {\bibfnamefont {M.~Achasov M.N.
  et~al.}\ \bibnamefont {The BESIII~Collaboration.}, \bibfnamefont {Ablikim}}
  (\bibinfo {collaboration} {BES}),\ }\bibfield  {title} {\enquote {\bibinfo
  {title} {{Polarization and Entanglement in Baryon-Antibaryon Pair Production
  in Electron-Positron Annihilation}},}\ }\href {\doibase
  10.1038/s41567-019-0494-8} {\bibfield  {journal} {\bibinfo  {journal} {Nat.
  Phys.}\ }\textbf {\bibinfo {volume} {15}},\ \bibinfo {pages} {631--634}
  (\bibinfo {year} {2019})}\BibitemShut {NoStop}%
\bibitem [{\citenamefont {Moortgat-Pick}\ \emph {et~al.}(2008)\citenamefont
  {Moortgat-Pick}, \citenamefont {Abe}, \citenamefont {Alexander},
  \citenamefont {Ananthanarayan}, \citenamefont {Babich}, \citenamefont
  {Bharadwaj}, \citenamefont {Barber}, \citenamefont {Bartl}, \citenamefont
  {Brachmann}, \citenamefont {Chen}, \citenamefont {Clarke}, \citenamefont
  {Clendenin}, \citenamefont {Dainton}, \citenamefont {Desch}, \citenamefont
  {Diehl}, \citenamefont {Dobos}, \citenamefont {Dorland}, \citenamefont
  {Dreiner}, \citenamefont {Eberl}, \citenamefont {Ellis}, \citenamefont
  {Flöttmann}, \citenamefont {Fraas}, \citenamefont {Franco-Sollova},
  \citenamefont {Franke}, \citenamefont {Freitas}, \citenamefont {Goodson},
  \citenamefont {Gray}, \citenamefont {Han}, \citenamefont {Heinemeyer},
  \citenamefont {Hesselbach}, \citenamefont {Hirose}, \citenamefont
  {Hohenwarter-Sodek}, \citenamefont {Juste}, \citenamefont {Kalinowski},
  \citenamefont {Kernreiter}, \citenamefont {Kittel}, \citenamefont {Kraml},
  \citenamefont {Langenfeld}, \citenamefont {Majerotto}, \citenamefont
  {Martinez}, \citenamefont {Martyn}, \citenamefont {Mikhailichenko},
  \citenamefont {Milstene}, \citenamefont {Menges}, \citenamefont {Meyners},
  \citenamefont {Mönig}, \citenamefont {Moffeit}, \citenamefont {Moretti},
  \citenamefont {Nachtmann}, \citenamefont {Nagel}, \citenamefont {Nakanishi},
  \citenamefont {Nauenberg}, \citenamefont {Nowak}, \citenamefont {Omori},
  \citenamefont {Osland}, \citenamefont {Pankov}, \citenamefont {Paver},
  \citenamefont {Pitthan}, \citenamefont {Pöschl}, \citenamefont {Porod},
  \citenamefont {Proulx}, \citenamefont {Richardson}, \citenamefont {Riemann},
  \citenamefont {Rindani}, \citenamefont {Rizzo}, \citenamefont {Schälicke},
  \citenamefont {Schüler}, \citenamefont {Schwanenberger}, \citenamefont
  {Scott}, \citenamefont {Sheppard}, \citenamefont {Singh}, \citenamefont
  {Sopczak}, \citenamefont {Spiesberger}, \citenamefont {Stahl}, \citenamefont
  {Steiner}, \citenamefont {Wagner}, \citenamefont {Weber}, \citenamefont
  {Weiglein}, \citenamefont {Wilson}, \citenamefont {Woods}, \citenamefont
  {Zerwas}, \citenamefont {Zhang},\ and\ \citenamefont {Zomer}}]{Moortgat2008}%
  \BibitemOpen
  \bibfield  {author} {\bibinfo {author} {\bibfnamefont {G.}~\bibnamefont
  {Moortgat-Pick}}, \bibinfo {author} {\bibfnamefont {T.}~\bibnamefont {Abe}},
  \bibinfo {author} {\bibfnamefont {G.}~\bibnamefont {Alexander}}, \bibinfo
  {author} {\bibfnamefont {B.}~\bibnamefont {Ananthanarayan}}, \bibinfo
  {author} {\bibfnamefont {A.A.}\ \bibnamefont {Babich}}, \bibinfo {author}
  {\bibfnamefont {V.}~\bibnamefont {Bharadwaj}}, \bibinfo {author}
  {\bibfnamefont {D.}~\bibnamefont {Barber}}, \bibinfo {author} {\bibfnamefont
  {A.}~\bibnamefont {Bartl}}, \bibinfo {author} {\bibfnamefont
  {A.}~\bibnamefont {Brachmann}}, \bibinfo {author} {\bibfnamefont
  {S.}~\bibnamefont {Chen}}, \bibinfo {author} {\bibfnamefont {J.}~\bibnamefont
  {Clarke}}, \bibinfo {author} {\bibfnamefont {J.E.}\ \bibnamefont
  {Clendenin}}, \bibinfo {author} {\bibfnamefont {J.}~\bibnamefont {Dainton}},
  \bibinfo {author} {\bibfnamefont {K.}~\bibnamefont {Desch}}, \bibinfo
  {author} {\bibfnamefont {M.}~\bibnamefont {Diehl}}, \bibinfo {author}
  {\bibfnamefont {B.}~\bibnamefont {Dobos}}, \bibinfo {author} {\bibfnamefont
  {T.}~\bibnamefont {Dorland}}, \bibinfo {author} {\bibfnamefont {H.K.}\
  \bibnamefont {Dreiner}}, \bibinfo {author} {\bibfnamefont {H.}~\bibnamefont
  {Eberl}}, \bibinfo {author} {\bibfnamefont {J.}~\bibnamefont {Ellis}},
  \bibinfo {author} {\bibfnamefont {K.}~\bibnamefont {Flöttmann}}, \bibinfo
  {author} {\bibfnamefont {H.}~\bibnamefont {Fraas}}, \bibinfo {author}
  {\bibfnamefont {F.}~\bibnamefont {Franco-Sollova}}, \bibinfo {author}
  {\bibfnamefont {F.}~\bibnamefont {Franke}}, \bibinfo {author} {\bibfnamefont
  {A.}~\bibnamefont {Freitas}}, \bibinfo {author} {\bibfnamefont
  {J.}~\bibnamefont {Goodson}}, \bibinfo {author} {\bibfnamefont
  {J.}~\bibnamefont {Gray}}, \bibinfo {author} {\bibfnamefont {A.}~\bibnamefont
  {Han}}, \bibinfo {author} {\bibfnamefont {S.}~\bibnamefont {Heinemeyer}},
  \bibinfo {author} {\bibfnamefont {S.}~\bibnamefont {Hesselbach}}, \bibinfo
  {author} {\bibfnamefont {T.}~\bibnamefont {Hirose}}, \bibinfo {author}
  {\bibfnamefont {K.}~\bibnamefont {Hohenwarter-Sodek}}, \bibinfo {author}
  {\bibfnamefont {A.}~\bibnamefont {Juste}}, \bibinfo {author} {\bibfnamefont
  {J.}~\bibnamefont {Kalinowski}}, \bibinfo {author} {\bibfnamefont
  {T.}~\bibnamefont {Kernreiter}}, \bibinfo {author} {\bibfnamefont
  {O.}~\bibnamefont {Kittel}}, \bibinfo {author} {\bibfnamefont
  {S.}~\bibnamefont {Kraml}}, \bibinfo {author} {\bibfnamefont
  {U.}~\bibnamefont {Langenfeld}}, \bibinfo {author} {\bibfnamefont
  {W.}~\bibnamefont {Majerotto}}, \bibinfo {author} {\bibfnamefont
  {A.}~\bibnamefont {Martinez}}, \bibinfo {author} {\bibfnamefont {H.-U.}\
  \bibnamefont {Martyn}}, \bibinfo {author} {\bibfnamefont {A.}~\bibnamefont
  {Mikhailichenko}}, \bibinfo {author} {\bibfnamefont {C.}~\bibnamefont
  {Milstene}}, \bibinfo {author} {\bibfnamefont {W.}~\bibnamefont {Menges}},
  \bibinfo {author} {\bibfnamefont {N.}~\bibnamefont {Meyners}}, \bibinfo
  {author} {\bibfnamefont {K.}~\bibnamefont {Mönig}}, \bibinfo {author}
  {\bibfnamefont {K.}~\bibnamefont {Moffeit}}, \bibinfo {author} {\bibfnamefont
  {S.}~\bibnamefont {Moretti}}, \bibinfo {author} {\bibfnamefont
  {O.}~\bibnamefont {Nachtmann}}, \bibinfo {author} {\bibfnamefont
  {F.}~\bibnamefont {Nagel}}, \bibinfo {author} {\bibfnamefont
  {T.}~\bibnamefont {Nakanishi}}, \bibinfo {author} {\bibfnamefont
  {U.}~\bibnamefont {Nauenberg}}, \bibinfo {author} {\bibfnamefont
  {H.}~\bibnamefont {Nowak}}, \bibinfo {author} {\bibfnamefont
  {T.}~\bibnamefont {Omori}}, \bibinfo {author} {\bibfnamefont
  {P.}~\bibnamefont {Osland}}, \bibinfo {author} {\bibfnamefont {A.A.}\
  \bibnamefont {Pankov}}, \bibinfo {author} {\bibfnamefont {N.}~\bibnamefont
  {Paver}}, \bibinfo {author} {\bibfnamefont {R.}~\bibnamefont {Pitthan}},
  \bibinfo {author} {\bibfnamefont {R.}~\bibnamefont {Pöschl}}, \bibinfo
  {author} {\bibfnamefont {W.}~\bibnamefont {Porod}}, \bibinfo {author}
  {\bibfnamefont {J.}~\bibnamefont {Proulx}}, \bibinfo {author} {\bibfnamefont
  {P.}~\bibnamefont {Richardson}}, \bibinfo {author} {\bibfnamefont
  {S.}~\bibnamefont {Riemann}}, \bibinfo {author} {\bibfnamefont {S.D.}\
  \bibnamefont {Rindani}}, \bibinfo {author} {\bibfnamefont {T.G.}\
  \bibnamefont {Rizzo}}, \bibinfo {author} {\bibfnamefont {A.}~\bibnamefont
  {Schälicke}}, \bibinfo {author} {\bibfnamefont {P.}~\bibnamefont
  {Schüler}}, \bibinfo {author} {\bibfnamefont {C.}~\bibnamefont
  {Schwanenberger}}, \bibinfo {author} {\bibfnamefont {D.}~\bibnamefont
  {Scott}}, \bibinfo {author} {\bibfnamefont {J.}~\bibnamefont {Sheppard}},
  \bibinfo {author} {\bibfnamefont {R.K.}\ \bibnamefont {Singh}}, \bibinfo
  {author} {\bibfnamefont {A.}~\bibnamefont {Sopczak}}, \bibinfo {author}
  {\bibfnamefont {H.}~\bibnamefont {Spiesberger}}, \bibinfo {author}
  {\bibfnamefont {A.}~\bibnamefont {Stahl}}, \bibinfo {author} {\bibfnamefont
  {H.}~\bibnamefont {Steiner}}, \bibinfo {author} {\bibfnamefont
  {A.}~\bibnamefont {Wagner}}, \bibinfo {author} {\bibfnamefont {A.M.}\
  \bibnamefont {Weber}}, \bibinfo {author} {\bibfnamefont {G.}~\bibnamefont
  {Weiglein}}, \bibinfo {author} {\bibfnamefont {G.W.}\ \bibnamefont {Wilson}},
  \bibinfo {author} {\bibfnamefont {M.}~\bibnamefont {Woods}}, \bibinfo
  {author} {\bibfnamefont {P.}~\bibnamefont {Zerwas}}, \bibinfo {author}
  {\bibfnamefont {J.}~\bibnamefont {Zhang}}, \ and\ \bibinfo {author}
  {\bibfnamefont {F.}~\bibnamefont {Zomer}},\ }\bibfield  {title} {\enquote
  {\bibinfo {title} {Polarized positrons and electrons at the linear
  collider},}\ }\href@noop {} {\bibfield  {journal} {\bibinfo  {journal} {Phys.
  Rep.}\ }\textbf {\bibinfo {volume} {460}},\ \bibinfo {pages} {131 -- 243}
  (\bibinfo {year} {2008})}\BibitemShut {NoStop}%
\bibitem [{\citenamefont {Bear}\ \emph {et~al.}(2013)\citenamefont {Bear},
  \citenamefont {Barklow}, \citenamefont {Fujii}, \citenamefont {Gao},
  \citenamefont {Hoang}, \citenamefont {Kanemura}, \citenamefont {List},
  \citenamefont {Logan}, \citenamefont {Nomerotski}, \citenamefont
  {Perelstein}, \citenamefont {Peskin}, \citenamefont {P\=oschl}, \citenamefont
  {Reuter}, \citenamefont {Riemann}, \citenamefont {Savoy-Navarro},
  \citenamefont {Servant}, \citenamefont {Tait},\ and\ \citenamefont
  {Yu}}]{baer2013}%
  \BibitemOpen
  \bibfield  {author} {\bibinfo {author} {\bibfnamefont {Howard}\ \bibnamefont
  {Bear}}, \bibinfo {author} {\bibfnamefont {Tim}\ \bibnamefont {Barklow}},
  \bibinfo {author} {\bibfnamefont {Keisuke}\ \bibnamefont {Fujii}}, \bibinfo
  {author} {\bibfnamefont {Yuanning}\ \bibnamefont {Gao}}, \bibinfo {author}
  {\bibfnamefont {Andre}\ \bibnamefont {Hoang}}, \bibinfo {author}
  {\bibfnamefont {Shinya}\ \bibnamefont {Kanemura}}, \bibinfo {author}
  {\bibfnamefont {Jenny}\ \bibnamefont {List}}, \bibinfo {author}
  {\bibfnamefont {Heather~E.}\ \bibnamefont {Logan}}, \bibinfo {author}
  {\bibfnamefont {Andrei}\ \bibnamefont {Nomerotski}}, \bibinfo {author}
  {\bibfnamefont {Maxim}\ \bibnamefont {Perelstein}}, \bibinfo {author}
  {\bibfnamefont {Michael~E.}\ \bibnamefont {Peskin}}, \bibinfo {author}
  {\bibfnamefont {Roman}\ \bibnamefont {P\=oschl}}, \bibinfo {author}
  {\bibfnamefont {J\"urgen}\ \bibnamefont {Reuter}}, \bibinfo {author}
  {\bibfnamefont {Sabine}\ \bibnamefont {Riemann}}, \bibinfo {author}
  {\bibfnamefont {Aurore}\ \bibnamefont {Savoy-Navarro}}, \bibinfo {author}
  {\bibfnamefont {Geraldine}\ \bibnamefont {Servant}}, \bibinfo {author}
  {\bibfnamefont {Tim M.~P.}\ \bibnamefont {Tait}}, \ and\ \bibinfo {author}
  {\bibfnamefont {Jaehoon}\ \bibnamefont {Yu}},\ }\href@noop {} {\enquote
  {\bibinfo {title} {The international linear collider technical design report
  - volume 2: Physics},}\ } (\bibinfo {year} {2013}),\ \Eprint
  {http://arxiv.org/abs/1306.6352} {arXiv:1306.6352 [hep-ph]} \BibitemShut
  {NoStop}%
\bibitem [{\citenamefont {Usun~Simitcioglu}\ \emph {et~al.}(2018)\citenamefont
  {Usun~Simitcioglu}, \citenamefont {Tapan}, \citenamefont {Tomas},
  \citenamefont {Barranco},\ and\ \citenamefont {Beckmann}}]{Simitcioglu2018}%
  \BibitemOpen
  \bibfield  {author} {\bibinfo {author} {\bibfnamefont {Aysegul}\ \bibnamefont
  {Usun~Simitcioglu}}, \bibinfo {author} {\bibfnamefont {Ilhan}\ \bibnamefont
  {Tapan}}, \bibinfo {author} {\bibfnamefont {R.}~\bibnamefont {Tomas}},
  \bibinfo {author} {\bibfnamefont {J.}~\bibnamefont {Barranco}}, \ and\
  \bibinfo {author} {\bibfnamefont {M.}~\bibnamefont {Beckmann}},\ }\bibfield
  {title} {\enquote {\bibinfo {title} {Beam offset impact on the polarization
  in the clic beam delivery system},}\ }\href {\doibase 10.1139/cjp-2017-0251}
  {\bibfield  {journal} {\bibinfo  {journal} {Can. J. Phys.}\ }\textbf
  {\bibinfo {volume} {96}},\ \bibinfo {pages} {1266--1271} (\bibinfo {year}
  {2018})}\BibitemShut {NoStop}%
\bibitem [{\citenamefont {Ari}\ \emph {et~al.}(2016)\citenamefont {Ari},
  \citenamefont {Billur}, \citenamefont {\.Inan},\ and\ \citenamefont
  {K\"oksal}}]{Ari2016}%
  \BibitemOpen
  \bibfield  {author} {\bibinfo {author} {\bibfnamefont {V.}~\bibnamefont
  {Ari}}, \bibinfo {author} {\bibfnamefont {A.A.}\ \bibnamefont {Billur}},
  \bibinfo {author} {\bibfnamefont {S.C.}\ \bibnamefont {\.Inan}}, \ and\
  \bibinfo {author} {\bibfnamefont {M.}~\bibnamefont {K\"oksal}},\ }\bibfield
  {title} {\enquote {\bibinfo {title} {Anomalous $ww\gamma$ couplings with beam
  polarization at the compact linear collider},}\ }\href {\doibase
  https://doi.org/10.1016/j.nuclphysb.2016.02.029} {\bibfield  {journal}
  {\bibinfo  {journal} {Nucl. Phys. B}\ }\textbf {\bibinfo {volume} {906}},\
  \bibinfo {pages} {211 -- 230} (\bibinfo {year} {2016})}\BibitemShut {NoStop}%
\bibitem [{\citenamefont {Duan}\ \emph {et~al.}(2019)\citenamefont {Duan},
  \citenamefont {Gao}, \citenamefont {Li}, \citenamefont {Wang}, \citenamefont
  {Wang}, \citenamefont {Xia}, \citenamefont {Xu}, \citenamefont {Yu},\ and\
  \citenamefont {Zhang}}]{Duan2019}%
  \BibitemOpen
  \bibfield  {author} {\bibinfo {author} {\bibfnamefont {Zhe}\ \bibnamefont
  {Duan}}, \bibinfo {author} {\bibfnamefont {Jie}\ \bibnamefont {Gao}},
  \bibinfo {author} {\bibfnamefont {Xiaoping}\ \bibnamefont {Li}}, \bibinfo
  {author} {\bibfnamefont {Dou}\ \bibnamefont {Wang}}, \bibinfo {author}
  {\bibfnamefont {Yiwei}\ \bibnamefont {Wang}}, \bibinfo {author}
  {\bibfnamefont {Wenhao}\ \bibnamefont {Xia}}, \bibinfo {author}
  {\bibfnamefont {Qingjin}\ \bibnamefont {Xu}}, \bibinfo {author}
  {\bibfnamefont {Chenghui}\ \bibnamefont {Yu}}, \ and\ \bibinfo {author}
  {\bibfnamefont {Yuan}\ \bibnamefont {Zhang}},\ }\bibfield  {title} {\enquote
  {\bibinfo {title} {Concepts of longitudinally polarized electron and positron
  colliding beams in the circular electron positron collider},}\ }in\ \href
  {\doibase 10.18429/JACoW-IPAC2019-MOPMP012} {\emph {\bibinfo {booktitle}
  {10th International Particle Accelerator Conference}}}\ (\bibinfo {year}
  {2019})\ p.\ \bibinfo {pages} {MOPMP012}\BibitemShut {NoStop}%
\bibitem [{\citenamefont {Nikitin}(2020)}]{Nikitin2020}%
  \BibitemOpen
  \bibfield  {author} {\bibinfo {author} {\bibfnamefont {Sergei}\ \bibnamefont
  {Nikitin}},\ }\bibfield  {title} {\enquote {\bibinfo {title} {Polarization
  issues in circular electron positron super-colliders},}\ }\href {\doibase
  10.1142/S0217751X20410018} {\bibfield  {journal} {\bibinfo  {journal} {Int.
  J. Mod. Phys. A}\ }\textbf {\bibinfo {volume} {35}},\ \bibinfo {pages}
  {2041001} (\bibinfo {year} {2020})}\BibitemShut {NoStop}%
\bibitem [{\citenamefont {Diehl}\ \emph {et~al.}(2003)\citenamefont {Diehl},
  \citenamefont {Nachtmann},\ and\ \citenamefont {Nagel}}]{Diehl2003}%
  \BibitemOpen
  \bibfield  {author} {\bibinfo {author} {\bibfnamefont {M.}~\bibnamefont
  {Diehl}}, \bibinfo {author} {\bibfnamefont {O.}~\bibnamefont {Nachtmann}}, \
  and\ \bibinfo {author} {\bibfnamefont {F.}~\bibnamefont {Nagel}},\ }\bibfield
   {title} {\enquote {\bibinfo {title} {{Probing triple gauge couplings with
  transverse beam polarisation in e +e- $\rightarrow$ W+W-}},}\ }\href
  {\doibase 10.1140/epjc/s2003-01339-5} {\bibfield  {journal} {\bibinfo
  {journal} {Eur. Phys. J. C}\ }\textbf {\bibinfo {volume} {32}},\ \bibinfo
  {pages} {17--27} (\bibinfo {year} {2003})}\BibitemShut {NoStop}%
\bibitem [{\citenamefont {Chakraborty}\ \emph {et~al.}(2003)\citenamefont
  {Chakraborty}, \citenamefont {Konigsberg},\ and\ \citenamefont
  {Rainwater}}]{chakraborty2003}%
  \BibitemOpen
  \bibfield  {author} {\bibinfo {author} {\bibfnamefont {Dhiman}\ \bibnamefont
  {Chakraborty}}, \bibinfo {author} {\bibfnamefont {Jacobo}\ \bibnamefont
  {Konigsberg}}, \ and\ \bibinfo {author} {\bibfnamefont {David}\ \bibnamefont
  {Rainwater}},\ }\bibfield  {title} {\enquote {\bibinfo {title} {Top-quark
  physics},}\ }\href {\doibase 10.1146/annurev.nucl.53.041002.110601}
  {\bibfield  {journal} {\bibinfo  {journal} {Annu. Rev. Nucl. Part. Sci.}\
  }\textbf {\bibinfo {volume} {53}},\ \bibinfo {pages} {301--351} (\bibinfo
  {year} {2003})}\BibitemShut {NoStop}%
\bibitem [{\citenamefont {Bartl}\ \emph {et~al.}(2007)\citenamefont {Bartl},
  \citenamefont {Hohenwarter-Sodek}, \citenamefont {Kernreiter},\ and\
  \citenamefont {Kittel}}]{Bartl2007}%
  \BibitemOpen
  \bibfield  {author} {\bibinfo {author} {\bibfnamefont {Alfred}\ \bibnamefont
  {Bartl}}, \bibinfo {author} {\bibfnamefont {Karl}\ \bibnamefont
  {Hohenwarter-Sodek}}, \bibinfo {author} {\bibfnamefont {Thomas}\ \bibnamefont
  {Kernreiter}}, \ and\ \bibinfo {author} {\bibfnamefont {Olaf}\ \bibnamefont
  {Kittel}},\ }\bibfield  {title} {\enquote {\bibinfo {title} {Cp asymmetries
  with longitudinal and transverse beam polarizations in neutralino production
  and decay into the z$^{0}$ boson at the ilc},}\ }\href@noop {} {\bibfield
  {journal} {\bibinfo  {journal} {J. High Energy Phys.}\ }\textbf {\bibinfo
  {volume} {2007}},\ \bibinfo {pages} {079--079} (\bibinfo {year}
  {2007})}\BibitemShut {NoStop}%
\bibitem [{\citenamefont {Herczeg}(2003)}]{Herczeg2003}%
  \BibitemOpen
  \bibfield  {author} {\bibinfo {author} {\bibfnamefont {Peter}\ \bibnamefont
  {Herczeg}},\ }\bibfield  {title} {\enquote {\bibinfo {title} {Cp-violating
  electron-nucleon interactions from leptoquark exchange},}\ }\href@noop {}
  {\bibfield  {journal} {\bibinfo  {journal} {Phys. Rev. D}\ }\textbf {\bibinfo
  {volume} {68}},\ \bibinfo {pages} {116004} (\bibinfo {year}
  {2003})}\BibitemShut {NoStop}%
\bibitem [{\citenamefont {Ananthanarayan}\ and\ \citenamefont
  {Rindani}(2004{\natexlab{a}})}]{Ananthanarayan2004}%
  \BibitemOpen
  \bibfield  {author} {\bibinfo {author} {\bibfnamefont {B.}~\bibnamefont
  {Ananthanarayan}}\ and\ \bibinfo {author} {\bibfnamefont {Saurabh~D.}\
  \bibnamefont {Rindani}},\ }\bibfield  {title} {\enquote {\bibinfo {title}
  {{CP violation at a linear collider with transverse polarization}},}\ }\href
  {\doibase 10.1103/PhysRevD.70.036005} {\bibfield  {journal} {\bibinfo
  {journal} {Phys. Rev. D}\ }\textbf {\bibinfo {volume} {70}},\ \bibinfo
  {pages} {036005} (\bibinfo {year} {2004}{\natexlab{a}})}\BibitemShut
  {NoStop}%
\bibitem [{\citenamefont {Ananthanarayan}\ and\ \citenamefont
  {Rindani}(2005)}]{Ananthanarayan_2005}%
  \BibitemOpen
  \bibfield  {author} {\bibinfo {author} {\bibfnamefont {B.}~\bibnamefont
  {Ananthanarayan}}\ and\ \bibinfo {author} {\bibfnamefont {Saurabh~D.}\
  \bibnamefont {Rindani}},\ }\bibfield  {title} {\enquote {\bibinfo {title}
  {Transverse beam polarization and cp violation in $e^+ e^- \rightarrow \gamma
  z$ with contact interactions},}\ }\href@noop {} {\bibfield  {journal}
  {\bibinfo  {journal} {Phys. Lett. B}\ }\textbf {\bibinfo {volume} {606}},\
  \bibinfo {pages} {107--115} (\bibinfo {year} {2005})}\BibitemShut {NoStop}%
\bibitem [{\citenamefont {Ananthanarayan}\ and\ \citenamefont
  {Rindani}(2018)}]{Ananthanarayan2018}%
  \BibitemOpen
  \bibfield  {author} {\bibinfo {author} {\bibfnamefont {B.}~\bibnamefont
  {Ananthanarayan}}\ and\ \bibinfo {author} {\bibfnamefont {Saurabh~D.}\
  \bibnamefont {Rindani}},\ }\bibfield  {title} {\enquote {\bibinfo {title}
  {{Inclusive spin-momentum analysis and new physics at a polarized
  electron-positron collider}},}\ }\href@noop {} {\bibfield  {journal}
  {\bibinfo  {journal} {Eur. Phys. J. C}\ }\textbf {\bibinfo {volume} {78}},\
  \bibinfo {pages} {1--17} (\bibinfo {year} {2018})}\BibitemShut {NoStop}%
\bibitem [{\citenamefont {Fleischer}\ \emph {et~al.}(1994)\citenamefont
  {Fleischer}, \citenamefont {Ko{\l}odziej},\ and\ \citenamefont
  {Jegerlehner}}]{Fleischer1994}%
  \BibitemOpen
  \bibfield  {author} {\bibinfo {author} {\bibfnamefont {J.}~\bibnamefont
  {Fleischer}}, \bibinfo {author} {\bibfnamefont {K.}~\bibnamefont
  {Ko{\l}odziej}}, \ and\ \bibinfo {author} {\bibfnamefont {F.}~\bibnamefont
  {Jegerlehner}},\ }\bibfield  {title} {\enquote {\bibinfo {title} {{Transverse
  versus longitudinal polarization effects in e+e- $\rightarrow$ +W-}},}\
  }\href@noop {} {\bibfield  {journal} {\bibinfo  {journal} {Phys. Rev. D}\
  }\textbf {\bibinfo {volume} {49}},\ \bibinfo {pages} {2174--2187} (\bibinfo
  {year} {1994})}\BibitemShut {NoStop}%
\bibitem [{\citenamefont {Hikasa}(1986)}]{Hikasa_1986}%
  \BibitemOpen
  \bibfield  {author} {\bibinfo {author} {\bibfnamefont {Ken-ichi}\
  \bibnamefont {Hikasa}},\ }\bibfield  {title} {\enquote {\bibinfo {title}
  {Transverse-polarization effects in ${e}^{+}$${e}^{\mathrm{\ensuremath{-}}}$
  collisions: The role of chiral symmetry},}\ }\href {\doibase
  10.1103/PhysRevD.33.3203} {\bibfield  {journal} {\bibinfo  {journal} {Phys.
  Rev. D}\ }\textbf {\bibinfo {volume} {33}},\ \bibinfo {pages} {3203--3223}
  (\bibinfo {year} {1986})}\BibitemShut {NoStop}%
\bibitem [{\citenamefont {Rizzo}(2003)}]{Rizzo2003}%
  \BibitemOpen
  \bibfield  {author} {\bibinfo {author} {\bibfnamefont {Thomas~G.}\
  \bibnamefont {Rizzo}},\ }\bibfield  {title} {\enquote {\bibinfo {title}
  {{Transverse polarization signatures of extra dimensions at linear
  colliders}},}\ }\href {\doibase 10.1088/1126-6708/2003/02/008} {\bibfield
  {journal} {\bibinfo  {journal} {J. High Energy Phys.}\ }\textbf {\bibinfo
  {volume} {7}},\ \bibinfo {pages} {157--172} (\bibinfo {year}
  {2003})}\BibitemShut {NoStop}%
\bibitem [{\citenamefont {Dass}\ and\ \citenamefont {Ross}(1977)}]{Dass1977}%
  \BibitemOpen
  \bibfield  {author} {\bibinfo {author} {\bibfnamefont {G.V.}\ \bibnamefont
  {Dass}}\ and\ \bibinfo {author} {\bibfnamefont {G.G.}\ \bibnamefont {Ross}},\
  }\bibfield  {title} {\enquote {\bibinfo {title} {Neutral and weak currents in
  $e^{+}e^{-}$ annihilation to hadrons},}\ }\href {\doibase
  https://doi.org/10.1016/0550-3213(77)90310-8} {\bibfield  {journal} {\bibinfo
   {journal} {Nucl. Phys. B}\ }\textbf {\bibinfo {volume} {118}},\ \bibinfo
  {pages} {284--310} (\bibinfo {year} {1977})}\BibitemShut {NoStop}%
\bibitem [{\citenamefont {Burgess}\ and\ \citenamefont
  {Robinson}(1991)}]{Burgess1991}%
  \BibitemOpen
  \bibfield  {author} {\bibinfo {author} {\bibfnamefont {C.~P.}\ \bibnamefont
  {Burgess}}\ and\ \bibinfo {author} {\bibfnamefont {J.~A.}\ \bibnamefont
  {Robinson}},\ }\bibfield  {title} {\enquote {\bibinfo {title} {{Transverse
  polarization at $e^+e^-$ colliders and CP violation from new physics}},}\
  }\href {\doibase 10.1142/S0217751X91001313} {\bibfield  {journal} {\bibinfo
  {journal} {Int. J. Mod. Phys. A}\ }\textbf {\bibinfo {volume} {6}},\ \bibinfo
  {pages} {2707--2728} (\bibinfo {year} {1991})}\BibitemShut {NoStop}%
\bibitem [{\citenamefont {Harlander}\ \emph {et~al.}(1997)\citenamefont
  {Harlander}, \citenamefont {Je\.zabek}, \citenamefont {K\"uhn},\ and\
  \citenamefont {Peter}}]{Harlander1997}%
  \BibitemOpen
  \bibfield  {author} {\bibinfo {author} {\bibfnamefont {R.}~\bibnamefont
  {Harlander}}, \bibinfo {author} {\bibfnamefont {M.}~\bibnamefont
  {Je\.zabek}}, \bibinfo {author} {\bibfnamefont {J.~H.}\ \bibnamefont
  {K\"uhn}}, \ and\ \bibinfo {author} {\bibfnamefont {M.}~\bibnamefont
  {Peter}},\ }\bibfield  {title} {\enquote {\bibinfo {title} {Top quark
  polarization in polarized $e^{+}e^{-}$ annihilation near threshold},}\ }\href
  {\doibase 10.1007/s002880050338} {\bibfield  {journal} {\bibinfo  {journal}
  {Z. Phys. C: Part. Fields}\ }\textbf {\bibinfo {volume} {73}},\ \bibinfo
  {pages} {477--494} (\bibinfo {year} {1997})}\BibitemShut {NoStop}%
\bibitem [{\citenamefont {Godbole}\ \emph {et~al.}(2006)\citenamefont
  {Godbole}, \citenamefont {Rindani},\ and\ \citenamefont
  {Singh}}]{Godbole_2006}%
  \BibitemOpen
  \bibfield  {author} {\bibinfo {author} {\bibfnamefont {Rohini~M}\
  \bibnamefont {Godbole}}, \bibinfo {author} {\bibfnamefont {Saurabh~D}\
  \bibnamefont {Rindani}}, \ and\ \bibinfo {author} {\bibfnamefont {Ritesh~K}\
  \bibnamefont {Singh}},\ }\bibfield  {title} {\enquote {\bibinfo {title}
  {Lepton distribution as a probe of new physics in production and decay of
  thetquark and its polarization},}\ }\href {\doibase
  10.1088/1126-6708/2006/12/021} {\bibfield  {journal} {\bibinfo  {journal} {J.
  High Energy Phys.}\ }\textbf {\bibinfo {volume} {2006}},\ \bibinfo {pages}
  {021--021} (\bibinfo {year} {2006})}\BibitemShut {NoStop}%
\bibitem [{\citenamefont {Choi}\ \emph {et~al.}(2015)\citenamefont {Choi},
  \citenamefont {Christensen}, \citenamefont {Salmon},\ and\ \citenamefont
  {Wang}}]{Choi2015}%
  \BibitemOpen
  \bibfield  {author} {\bibinfo {author} {\bibfnamefont {S.~Y.}\ \bibnamefont
  {Choi}}, \bibinfo {author} {\bibfnamefont {N.~D.}\ \bibnamefont
  {Christensen}}, \bibinfo {author} {\bibfnamefont {D.}~\bibnamefont {Salmon}},
  \ and\ \bibinfo {author} {\bibfnamefont {X.}~\bibnamefont {Wang}},\
  }\bibfield  {title} {\enquote {\bibinfo {title} {{Spin and chirality effects
  in antler-topology processes at high energy $e^{+}e^{-}$ colliders}},}\
  }\href@noop {} {\bibfield  {journal} {\bibinfo  {journal} {Eur. Phys. J. C}\
  }\textbf {\bibinfo {volume} {75}},\ \bibinfo {pages} {481} (\bibinfo {year}
  {2015})}\BibitemShut {NoStop}%
\bibitem [{\citenamefont {Sokolov}\ and\ \citenamefont
  {Ternov}(1964)}]{Sokolov_1964}%
  \BibitemOpen
  \bibfield  {author} {\bibinfo {author} {\bibfnamefont {A.~A.}\ \bibnamefont
  {Sokolov}}\ and\ \bibinfo {author} {\bibfnamefont {I.~M.}\ \bibnamefont
  {Ternov}},\ }\bibfield  {title} {\enquote {\bibinfo {title} {On polarization
  and spin effects in the theory of synchrotron radiation},}\ }\href@noop {}
  {\bibfield  {journal} {\bibinfo  {journal} {Sov. Phys. Dokl.}\ }\textbf
  {\bibinfo {volume} {8}},\ \bibinfo {pages} {1203} (\bibinfo {year}
  {1964})}\BibitemShut {NoStop}%
\bibitem [{\citenamefont {Sokolov}\ and\ \citenamefont
  {Ternov}(1968)}]{Sokolov_1968}%
  \BibitemOpen
  \bibfield  {author} {\bibinfo {author} {\bibfnamefont {A.~A.}\ \bibnamefont
  {Sokolov}}\ and\ \bibinfo {author} {\bibfnamefont {I.~M.}\ \bibnamefont
  {Ternov}},\ }\href@noop {} {\emph {\bibinfo {title} {Synchrotron
  Radiation}}}\ (\bibinfo  {publisher} {Akademic, Germany},\ \bibinfo {year}
  {1968})\BibitemShut {NoStop}%
\bibitem [{\citenamefont {Baier}\ and\ \citenamefont
  {Katkov}(1967)}]{Baier_1967}%
  \BibitemOpen
  \bibfield  {author} {\bibinfo {author} {\bibfnamefont {V.~N.}\ \bibnamefont
  {Baier}}\ and\ \bibinfo {author} {\bibfnamefont {V.~M.}\ \bibnamefont
  {Katkov}},\ }\bibfield  {title} {\enquote {\bibinfo {title} {{Radiational
  polarization of electrons in inhomogeneous magnetic field}},}\ }\href@noop {}
  {\bibfield  {journal} {\bibinfo  {journal} {Phys. Lett. A}\ }\textbf
  {\bibinfo {volume} {24}},\ \bibinfo {pages} {327--329} (\bibinfo {year}
  {1967})}\BibitemShut {NoStop}%
\bibitem [{\citenamefont {Baier}(1972)}]{Baier_1972}%
  \BibitemOpen
  \bibfield  {author} {\bibinfo {author} {\bibfnamefont {V.~N.}\ \bibnamefont
  {Baier}},\ }\bibfield  {title} {\enquote {\bibinfo {title} {Radiative
  polarization of electron in storage rings},}\ }\href@noop {} {\bibfield
  {journal} {\bibinfo  {journal} {Sov. Phys. Usp.}\ }\textbf {\bibinfo {volume}
  {14}},\ \bibinfo {pages} {695} (\bibinfo {year} {1972})}\BibitemShut
  {NoStop}%
\bibitem [{\citenamefont {Derbenev}\ and\ \citenamefont
  {Kondratenko}(1973)}]{Derbenev_1973}%
  \BibitemOpen
  \bibfield  {author} {\bibinfo {author} {\bibfnamefont {Y.}~\bibnamefont
  {Derbenev}}\ and\ \bibinfo {author} {\bibfnamefont {A.~M.}\ \bibnamefont
  {Kondratenko}},\ }\bibfield  {title} {\enquote {\bibinfo {title}
  {{Polarization kinematics of particles in storage rings}},}\ }\href@noop {}
  {\bibfield  {journal} {\bibinfo  {journal} {Sov. Phys. JETP}\ }\textbf
  {\bibinfo {volume} {37}},\ \bibinfo {pages} {968} (\bibinfo {year}
  {1973})}\BibitemShut {NoStop}%
\bibitem [{\citenamefont {Omori}\ \emph {et~al.}(2006)\citenamefont {Omori},
  \citenamefont {Fukuda}, \citenamefont {Hirose}, \citenamefont {Kurihara},
  \citenamefont {Kuroda}, \citenamefont {Nomura}, \citenamefont {Ohashi},
  \citenamefont {Okugi}, \citenamefont {Sakaue}, \citenamefont {Saito},
  \citenamefont {Urakawa}, \citenamefont {Washio},\ and\ \citenamefont
  {Yamazaki}}]{Omori_2006}%
  \BibitemOpen
  \bibfield  {author} {\bibinfo {author} {\bibfnamefont {T.}~\bibnamefont
  {Omori}}, \bibinfo {author} {\bibfnamefont {M.}~\bibnamefont {Fukuda}},
  \bibinfo {author} {\bibfnamefont {T.}~\bibnamefont {Hirose}}, \bibinfo
  {author} {\bibfnamefont {Y.}~\bibnamefont {Kurihara}}, \bibinfo {author}
  {\bibfnamefont {R.}~\bibnamefont {Kuroda}}, \bibinfo {author} {\bibfnamefont
  {M.}~\bibnamefont {Nomura}}, \bibinfo {author} {\bibfnamefont
  {A.}~\bibnamefont {Ohashi}}, \bibinfo {author} {\bibfnamefont
  {T.}~\bibnamefont {Okugi}}, \bibinfo {author} {\bibfnamefont
  {K.}~\bibnamefont {Sakaue}}, \bibinfo {author} {\bibfnamefont
  {T.}~\bibnamefont {Saito}}, \bibinfo {author} {\bibfnamefont
  {J.}~\bibnamefont {Urakawa}}, \bibinfo {author} {\bibfnamefont
  {M.}~\bibnamefont {Washio}}, \ and\ \bibinfo {author} {\bibfnamefont
  {I.}~\bibnamefont {Yamazaki}},\ }\bibfield  {title} {\enquote {\bibinfo
  {title} {Efficient propagation of polarization from laser photons to
  positrons through compton scattering and electron-positron pair creation},}\
  }\href@noop {} {\bibfield  {journal} {\bibinfo  {journal} {Phys. Rev. Lett.}\
  }\textbf {\bibinfo {volume} {96}},\ \bibinfo {pages} {114801} (\bibinfo
  {year} {2006})}\BibitemShut {NoStop}%
\bibitem [{\citenamefont {Alexander}\ \emph {et~al.}(2008)\citenamefont
  {Alexander}, \citenamefont {Barley}, \citenamefont {Batygin}, \citenamefont
  {Berridge}, \citenamefont {Bharadwaj}, \citenamefont {Bower}, \citenamefont
  {Bugg}, \citenamefont {Decker}, \citenamefont {Dollan}, \citenamefont
  {Efremenko}, \citenamefont {Gharibyan}, \citenamefont {Hast}, \citenamefont
  {Iverson}, \citenamefont {Kolanoski}, \citenamefont {Kovermann},
  \citenamefont {Laihem}, \citenamefont {Lohse}, \citenamefont {McDonald},
  \citenamefont {Mikhailichenko}, \citenamefont {Moortgat-Pick}, \citenamefont
  {Pahl}, \citenamefont {Pitthan}, \citenamefont {P\"oschl}, \citenamefont
  {Reinherz-Aronis}, \citenamefont {Riemann}, \citenamefont {Sch\"alicke},
  \citenamefont {Sch\"uler}, \citenamefont {Schweizer}, \citenamefont {Scott},
  \citenamefont {Sheppard}, \citenamefont {Stahl}, \citenamefont {Szalata},
  \citenamefont {Walz},\ and\ \citenamefont {Weidemann}}]{Alexander_2008}%
  \BibitemOpen
  \bibfield  {author} {\bibinfo {author} {\bibfnamefont {G.}~\bibnamefont
  {Alexander}}, \bibinfo {author} {\bibfnamefont {J.}~\bibnamefont {Barley}},
  \bibinfo {author} {\bibfnamefont {Y.}~\bibnamefont {Batygin}}, \bibinfo
  {author} {\bibfnamefont {S.}~\bibnamefont {Berridge}}, \bibinfo {author}
  {\bibfnamefont {V.}~\bibnamefont {Bharadwaj}}, \bibinfo {author}
  {\bibfnamefont {G.}~\bibnamefont {Bower}}, \bibinfo {author} {\bibfnamefont
  {W.}~\bibnamefont {Bugg}}, \bibinfo {author} {\bibfnamefont {F.-J.}\
  \bibnamefont {Decker}}, \bibinfo {author} {\bibfnamefont {R.}~\bibnamefont
  {Dollan}}, \bibinfo {author} {\bibfnamefont {Y.}~\bibnamefont {Efremenko}},
  \bibinfo {author} {\bibfnamefont {V.}~\bibnamefont {Gharibyan}}, \bibinfo
  {author} {\bibfnamefont {C.}~\bibnamefont {Hast}}, \bibinfo {author}
  {\bibfnamefont {R.}~\bibnamefont {Iverson}}, \bibinfo {author} {\bibfnamefont
  {H.}~\bibnamefont {Kolanoski}}, \bibinfo {author} {\bibfnamefont
  {J.}~\bibnamefont {Kovermann}}, \bibinfo {author} {\bibfnamefont
  {K.}~\bibnamefont {Laihem}}, \bibinfo {author} {\bibfnamefont
  {T.}~\bibnamefont {Lohse}}, \bibinfo {author} {\bibfnamefont {K.~T.}\
  \bibnamefont {McDonald}}, \bibinfo {author} {\bibfnamefont {A.~A.}\
  \bibnamefont {Mikhailichenko}}, \bibinfo {author} {\bibfnamefont {G.~A.}\
  \bibnamefont {Moortgat-Pick}}, \bibinfo {author} {\bibfnamefont
  {P.}~\bibnamefont {Pahl}}, \bibinfo {author} {\bibfnamefont {R.}~\bibnamefont
  {Pitthan}}, \bibinfo {author} {\bibfnamefont {R.}~\bibnamefont {P\"oschl}},
  \bibinfo {author} {\bibfnamefont {E.}~\bibnamefont {Reinherz-Aronis}},
  \bibinfo {author} {\bibfnamefont {S.}~\bibnamefont {Riemann}}, \bibinfo
  {author} {\bibfnamefont {A.}~\bibnamefont {Sch\"alicke}}, \bibinfo {author}
  {\bibfnamefont {K.~P.}\ \bibnamefont {Sch\"uler}}, \bibinfo {author}
  {\bibfnamefont {T.}~\bibnamefont {Schweizer}}, \bibinfo {author}
  {\bibfnamefont {D.}~\bibnamefont {Scott}}, \bibinfo {author} {\bibfnamefont
  {J.~C.}\ \bibnamefont {Sheppard}}, \bibinfo {author} {\bibfnamefont
  {A.}~\bibnamefont {Stahl}}, \bibinfo {author} {\bibfnamefont {Z.~M.}\
  \bibnamefont {Szalata}}, \bibinfo {author} {\bibfnamefont {D.}~\bibnamefont
  {Walz}}, \ and\ \bibinfo {author} {\bibfnamefont {A.~W.}\ \bibnamefont
  {Weidemann}},\ }\bibfield  {title} {\enquote {\bibinfo {title} {Observation
  of polarized positrons from an undulator-based source},}\ }\href@noop {}
  {\bibfield  {journal} {\bibinfo  {journal} {Phys. Rev. Lett.}\ }\textbf
  {\bibinfo {volume} {100}},\ \bibinfo {pages} {210801} (\bibinfo {year}
  {2008})}\BibitemShut {NoStop}%
\bibitem [{\citenamefont {Abbott}\ \emph {et~al.}(2016)\citenamefont {Abbott},
  \citenamefont {Adderley}, \citenamefont {Adeyemi}, \citenamefont {Aguilera},
  \citenamefont {Ali} \emph {et~al.}}]{abbott2016prl}%
  \BibitemOpen
  \bibfield  {author} {\bibinfo {author} {\bibfnamefont {D.}~\bibnamefont
  {Abbott}}, \bibinfo {author} {\bibfnamefont {P.}~\bibnamefont {Adderley}},
  \bibinfo {author} {\bibfnamefont {A.}~\bibnamefont {Adeyemi}}, \bibinfo
  {author} {\bibfnamefont {P.}~\bibnamefont {Aguilera}}, \bibinfo {author}
  {\bibfnamefont {M.}~\bibnamefont {Ali}},  \emph {et~al.} (\bibinfo
  {collaboration} {PEPPo Collaboration}),\ }\bibfield  {title} {\enquote
  {\bibinfo {title} {Production of highly polarized positrons using polarized
  electrons at mev energies},}\ }\href {\doibase
  10.1103/PhysRevLett.116.214801} {\bibfield  {journal} {\bibinfo  {journal}
  {Phys. Rev. Lett.}\ }\textbf {\bibinfo {volume} {116}},\ \bibinfo {pages}
  {214801} (\bibinfo {year} {2016})}\BibitemShut {NoStop}%
\bibitem [{\citenamefont {Heitler}(1954)}]{Heitler_1954}%
  \BibitemOpen
  \bibfield  {author} {\bibinfo {author} {\bibfnamefont {W.}~\bibnamefont
  {Heitler}},\ }\href@noop {} {\emph {\bibinfo {title} {The Quantum Theory of
  Radiation}}}\ (\bibinfo  {publisher} {Clarendon Press, Oxford},\ \bibinfo
  {year} {1954})\BibitemShut {NoStop}%
\bibitem [{\citenamefont {Variola}(2014)}]{Variola_2014}%
  \BibitemOpen
  \bibfield  {author} {\bibinfo {author} {\bibfnamefont {A.}~\bibnamefont
  {Variola}},\ }\bibfield  {title} {\enquote {\bibinfo {title} {{Advanced
  positron sources}},}\ }\href@noop {} {\bibfield  {journal} {\bibinfo
  {journal} {Nucl. Instr. Meth. Phys. Res. A}\ }\textbf {\bibinfo {volume}
  {740}},\ \bibinfo {pages} {21--26} (\bibinfo {year} {2014})}\BibitemShut
  {NoStop}%
\bibitem [{\citenamefont {Buon}\ and\ \citenamefont
  {Steffen}(1986)}]{Buon_1986}%
  \BibitemOpen
  \bibfield  {author} {\bibinfo {author} {\bibfnamefont {Jean}\ \bibnamefont
  {Buon}}\ and\ \bibinfo {author} {\bibfnamefont {Klaus}\ \bibnamefont
  {Steffen}},\ }\bibfield  {title} {\enquote {\bibinfo {title} {{Hera
  variable-energy "mini" spin rotator and head-on ep collision scheme with
  choice of electron helicity}},}\ }\href@noop {} {\bibfield  {journal}
  {\bibinfo  {journal} {Nucl. Instrum. Methods Phys. Res., Sect. A}\ }\textbf
  {\bibinfo {volume} {245}},\ \bibinfo {pages} {248--261} (\bibinfo {year}
  {1986})}\BibitemShut {NoStop}%
\bibitem [{\citenamefont {Moffeit}\ \emph {et~al.}(2005)\citenamefont
  {Moffeit}, \citenamefont {Woods}, \citenamefont {Sch{\"u}ler}, \citenamefont
  {M{\"o}nig},\ and\ \citenamefont {Bambade}}]{Moffeit_2005}%
  \BibitemOpen
  \bibfield  {author} {\bibinfo {author} {\bibfnamefont {K.}~\bibnamefont
  {Moffeit}}, \bibinfo {author} {\bibfnamefont {M.}~\bibnamefont {Woods}},
  \bibinfo {author} {\bibfnamefont {P.}~\bibnamefont {Sch{\"u}ler}}, \bibinfo
  {author} {\bibfnamefont {K.}~\bibnamefont {M{\"o}nig}}, \ and\ \bibinfo
  {author} {\bibfnamefont {P.}~\bibnamefont {Bambade}},\ }\bibfield  {title}
  {\enquote {\bibinfo {title} {Spin rotation schemes at the ilc for two
  interaction regions and positron polarization with both helicities},}\
  }\href@noop {} {\bibfield  {journal} {\bibinfo  {journal} {SLAC-TN-05-045,
  LCC-0159, IPBI-TN-2005-2}\ } (\bibinfo {year} {2005})}\BibitemShut {NoStop}%
\bibitem [{\citenamefont {Yoon}\ \emph {et~al.}(2019)\citenamefont {Yoon},
  \citenamefont {Jeon}, \citenamefont {Shin}, \citenamefont {Lee},
  \citenamefont {Lee}, \citenamefont {Choi}, \citenamefont {Kim}, \citenamefont
  {Sung},\ and\ \citenamefont {Nam}}]{Yoon2019}%
  \BibitemOpen
  \bibfield  {author} {\bibinfo {author} {\bibfnamefont {Jin~Woo}\ \bibnamefont
  {Yoon}}, \bibinfo {author} {\bibfnamefont {Cheonha}\ \bibnamefont {Jeon}},
  \bibinfo {author} {\bibfnamefont {Junghoon}\ \bibnamefont {Shin}}, \bibinfo
  {author} {\bibfnamefont {Seong~Ku}\ \bibnamefont {Lee}}, \bibinfo {author}
  {\bibfnamefont {Hwang~Woon}\ \bibnamefont {Lee}}, \bibinfo {author}
  {\bibfnamefont {Il~Woo}\ \bibnamefont {Choi}}, \bibinfo {author}
  {\bibfnamefont {Hyung~Taek}\ \bibnamefont {Kim}}, \bibinfo {author}
  {\bibfnamefont {Jae~Hee}\ \bibnamefont {Sung}}, \ and\ \bibinfo {author}
  {\bibfnamefont {Chang~Hee}\ \bibnamefont {Nam}},\ }\bibfield  {title}
  {\enquote {\bibinfo {title} {Achieving the laser intensity of $5.5\times
  10^{22}$ w/cm$^2$ with a wavefront-corrected multi-pw laser},}\ }\href
  {\doibase 10.1364/OE.27.020412} {\bibfield  {journal} {\bibinfo  {journal}
  {Opt. Express}\ }\textbf {\bibinfo {volume} {27}},\ \bibinfo {pages}
  {20412--20420} (\bibinfo {year} {2019})}\BibitemShut {NoStop}%
\bibitem [{\citenamefont {Danson}\ \emph {et~al.}(2019)\citenamefont {Danson},
  \citenamefont {Haefner}, \citenamefont {Bromage}, \citenamefont {Butcher},
  \citenamefont {Chanteloup}, \citenamefont {Chowdhury}, \citenamefont
  {Galvanauskas}, \citenamefont {Gizzi}, \citenamefont {Hein}, \citenamefont
  {Hillier},\ and\ \citenamefont {et~al.}}]{Danson_2019}%
  \BibitemOpen
  \bibfield  {author} {\bibinfo {author} {\bibfnamefont {Colin~N.}\
  \bibnamefont {Danson}}, \bibinfo {author} {\bibfnamefont {Constantin}\
  \bibnamefont {Haefner}}, \bibinfo {author} {\bibfnamefont {Jake}\
  \bibnamefont {Bromage}}, \bibinfo {author} {\bibfnamefont {Thomas}\
  \bibnamefont {Butcher}}, \bibinfo {author} {\bibfnamefont
  {Jean-Christophe~F.}\ \bibnamefont {Chanteloup}}, \bibinfo {author}
  {\bibfnamefont {Enam~A.}\ \bibnamefont {Chowdhury}}, \bibinfo {author}
  {\bibfnamefont {Almantas}\ \bibnamefont {Galvanauskas}}, \bibinfo {author}
  {\bibfnamefont {Leonida~A.}\ \bibnamefont {Gizzi}}, \bibinfo {author}
  {\bibfnamefont {Joachim}\ \bibnamefont {Hein}}, \bibinfo {author}
  {\bibfnamefont {David~I.}\ \bibnamefont {Hillier}}, \ and\ \bibinfo {author}
  {\bibnamefont {et~al.}},\ }\bibfield  {title} {\enquote {\bibinfo {title}
  {Petawatt and exawatt class lasers worldwide},}\ }\href {\doibase
  10.1017/hpl.2019.36} {\bibfield  {journal} {\bibinfo  {journal} {High Power
  Laser Sci. Eng.}\ }\textbf {\bibinfo {volume} {7}},\ \bibinfo {pages} {e54}
  (\bibinfo {year} {2019})}\BibitemShut {NoStop}%
\bibitem [{\citenamefont {Gales}\ \emph {et~al.}(2018)\citenamefont {Gales},
  \citenamefont {Tanaka}, \citenamefont {Balabanski}, \citenamefont {Negoita},
  \citenamefont {Stutman}, \citenamefont {Tesileanu}, \citenamefont {Ur},
  \citenamefont {Ursescu}, \citenamefont {An-drei}, \citenamefont {Ataman},
  \citenamefont {Cernaianu}, \citenamefont {DAlessi}, \citenamefont {Dancus},
  \citenamefont {Diaconescu}, \citenamefont {Djourelov}, \citenamefont
  {Filipescu}, \citenamefont {Ghenuche}, \citenamefont {Ghita}, \citenamefont
  {Matei}, \citenamefont {Seto}, \citenamefont {Zeng},\ and\ \citenamefont
  {Zamfir}}]{Gales_2018}%
  \BibitemOpen
  \bibfield  {author} {\bibinfo {author} {\bibfnamefont {S.}~\bibnamefont
  {Gales}}, \bibinfo {author} {\bibfnamefont {K.~A.}\ \bibnamefont {Tanaka}},
  \bibinfo {author} {\bibfnamefont {D.~L.}\ \bibnamefont {Balabanski}},
  \bibinfo {author} {\bibfnamefont {F.}~\bibnamefont {Negoita}}, \bibinfo
  {author} {\bibfnamefont {D.}~\bibnamefont {Stutman}}, \bibinfo {author}
  {\bibfnamefont {O.}~\bibnamefont {Tesileanu}}, \bibinfo {author}
  {\bibfnamefont {C.~A.}\ \bibnamefont {Ur}}, \bibinfo {author} {\bibfnamefont
  {D.}~\bibnamefont {Ursescu}}, \bibinfo {author} {\bibfnamefont
  {I.}~\bibnamefont {An-drei}}, \bibinfo {author} {\bibfnamefont
  {S.}~\bibnamefont {Ataman}}, \bibinfo {author} {\bibfnamefont {M.~O.}\
  \bibnamefont {Cernaianu}}, \bibinfo {author} {\bibfnamefont {L.}~\bibnamefont
  {DAlessi}}, \bibinfo {author} {\bibfnamefont {I.}~\bibnamefont {Dancus}},
  \bibinfo {author} {\bibfnamefont {B.}~\bibnamefont {Diaconescu}}, \bibinfo
  {author} {\bibfnamefont {N.}~\bibnamefont {Djourelov}}, \bibinfo {author}
  {\bibfnamefont {D.}~\bibnamefont {Filipescu}}, \bibinfo {author}
  {\bibfnamefont {P.}~\bibnamefont {Ghenuche}}, \bibinfo {author}
  {\bibfnamefont {D.~G.}\ \bibnamefont {Ghita}}, \bibinfo {author}
  {\bibfnamefont {C.}~\bibnamefont {Matei}}, \bibinfo {author} {\bibfnamefont
  {K.}~\bibnamefont {Seto}}, \bibinfo {author} {\bibfnamefont {M.}~\bibnamefont
  {Zeng}}, \ and\ \bibinfo {author} {\bibfnamefont {N.~V.}\ \bibnamefont
  {Zamfir}},\ }\bibfield  {title} {\enquote {\bibinfo {title} {The extreme
  light infrastructure nuclear physics (eli-np) facility: new horizons in
  physics with 10 pw ultra-intense lasers and 20 mev brilliant gamma beams},}\
  }\href@noop {} {\bibfield  {journal} {\bibinfo  {journal} {Rep. Prog. Phys.}\
  }\textbf {\bibinfo {volume} {81}},\ \bibinfo {pages} {094301} (\bibinfo
  {year} {2018})}\BibitemShut {NoStop}%
\bibitem [{\citenamefont {Ritus}(1985)}]{Ritus_1985}%
  \BibitemOpen
  \bibfield  {author} {\bibinfo {author} {\bibfnamefont {V.~I.}\ \bibnamefont
  {Ritus}},\ }\bibfield  {title} {\enquote {\bibinfo {title} {Quantum effects
  of the interaction of elementary particles with an intense electromagnetic
  field},}\ }\href@noop {} {\bibfield  {journal} {\bibinfo  {journal} {J. Sov.
  Laser Res.}\ }\textbf {\bibinfo {volume} {6}},\ \bibinfo {pages} {497}
  (\bibinfo {year} {1985})}\BibitemShut {NoStop}%
\bibitem [{\citenamefont {Goldman}(1964)}]{Goldman_1964}%
  \BibitemOpen
  \bibfield  {author} {\bibinfo {author} {\bibfnamefont {I.~I.}\ \bibnamefont
  {Goldman}},\ }\bibfield  {title} {\enquote {\bibinfo {title} {Intensity
  effects in compton scattering},}\ }\href@noop {} {\bibfield  {journal}
  {\bibinfo  {journal} {Sov. Phys. JETP}\ }\textbf {\bibinfo {volume} {19}},\
  \bibinfo {pages} {954} (\bibinfo {year} {1964})},\ \bibinfo {note} {{[Zh.
  Eksp. Teor. Fiz. 46, 1412 (1964)]}}\BibitemShut {NoStop}%
\bibitem [{\citenamefont {Nikishov}\ and\ \citenamefont
  {Ritus}(1964)}]{Nikishov_1964}%
  \BibitemOpen
  \bibfield  {author} {\bibinfo {author} {\bibfnamefont {A.~I.}\ \bibnamefont
  {Nikishov}}\ and\ \bibinfo {author} {\bibfnamefont {V.~I.}\ \bibnamefont
  {Ritus}},\ }\bibfield  {title} {\enquote {\bibinfo {title} {Quantum processes
  in the field of a plane electromagnetic wave and in a constant field. i},}\
  }\href@noop {} {\bibfield  {journal} {\bibinfo  {journal} {Sov. Phys. JETP}\
  }\textbf {\bibinfo {volume} {19}},\ \bibinfo {pages} {529} (\bibinfo {year}
  {1964})},\ \bibinfo {note} {{[Zh. Eksp. Teor. Fiz. 46, 776
  (1964)]}}\BibitemShut {NoStop}%
\bibitem [{\citenamefont {Brown}\ and\ \citenamefont
  {Kibble}(1964)}]{Brown_1964}%
  \BibitemOpen
  \bibfield  {author} {\bibinfo {author} {\bibfnamefont {Lowell~S.}\
  \bibnamefont {Brown}}\ and\ \bibinfo {author} {\bibfnamefont {T.~W.~B.}\
  \bibnamefont {Kibble}},\ }\bibfield  {title} {\enquote {\bibinfo {title}
  {Interaction of intense laser beams with electrons},}\ }\href {\doibase
  10.1103/PhysRev.133.A705} {\bibfield  {journal} {\bibinfo  {journal} {Phys.
  Rev.}\ }\textbf {\bibinfo {volume} {133}},\ \bibinfo {pages} {A705--A719}
  (\bibinfo {year} {1964})}\BibitemShut {NoStop}%
\bibitem [{CAI()}]{CAIN}%
  \BibitemOpen
  \href@noop {} {}\bibinfo {howpublished} {K. Yokoya, CAIN2.42 Users Manual.,
  \url{https://ilc.kek.jp/~yokoya/CAIN/Cain242/}}\BibitemShut {NoStop}%
\bibitem [{\citenamefont {Reiss}(1962)}]{Reiss1962}%
  \BibitemOpen
  \bibfield  {author} {\bibinfo {author} {\bibfnamefont {Howard~R.}\
  \bibnamefont {Reiss}},\ }\bibfield  {title} {\enquote {\bibinfo {title}
  {Absorption of light by light},}\ }\href {\doibase 10.1063/1.1703787}
  {\bibfield  {journal} {\bibinfo  {journal} {J. Math. Phys.}\ }\textbf
  {\bibinfo {volume} {3}},\ \bibinfo {pages} {59--67} (\bibinfo {year}
  {1962})}\BibitemShut {NoStop}%
\bibitem [{\citenamefont {Ivanov}\ \emph {et~al.}(2005)\citenamefont {Ivanov},
  \citenamefont {Kotkin},\ and\ \citenamefont {Serbo}}]{Ivanov_2005}%
  \BibitemOpen
  \bibfield  {author} {\bibinfo {author} {\bibfnamefont {D.~Y.}\ \bibnamefont
  {Ivanov}}, \bibinfo {author} {\bibfnamefont {G.~L.}\ \bibnamefont {Kotkin}},
  \ and\ \bibinfo {author} {\bibfnamefont {V.~G.}\ \bibnamefont {Serbo}},\
  }\bibfield  {title} {\enquote {\bibinfo {title} {{Complete description of
  polarization effects in $e^+ e^-$ pair production by a photon in the field of
  a strong laser wave}},}\ }\href@noop {} {\bibfield  {journal} {\bibinfo
  {journal} {Eur. Phys. J. C}\ }\textbf {\bibinfo {volume} {40}},\ \bibinfo
  {pages} {27} (\bibinfo {year} {2005})}\BibitemShut {NoStop}%
\bibitem [{\citenamefont {Seipt}\ and\ \citenamefont
  {King}(2020)}]{Seipt_2020}%
  \BibitemOpen
  \bibfield  {author} {\bibinfo {author} {\bibfnamefont {D.}~\bibnamefont
  {Seipt}}\ and\ \bibinfo {author} {\bibfnamefont {B.}~\bibnamefont {King}},\
  }\bibfield  {title} {\enquote {\bibinfo {title} {{Spin- and
  polarization-dependent locally-constant-field-approximation rates for
  nonlinear Compton and Breit-Wheeler processes}},}\ }\href {\doibase
  10.1103/PhysRevA.102.052805} {\bibfield  {journal} {\bibinfo  {journal}
  {Phys. Rev. A}\ }\textbf {\bibinfo {volume} {102}},\ \bibinfo {pages}
  {052805} (\bibinfo {year} {2020})}\BibitemShut {NoStop}%
\bibitem [{\citenamefont {Wistisen}(2020)}]{Wistisen_2020}%
  \BibitemOpen
  \bibfield  {author} {\bibinfo {author} {\bibfnamefont {Tobias~N.}\
  \bibnamefont {Wistisen}},\ }\bibfield  {title} {\enquote {\bibinfo {title}
  {Numerical approach to the semiclassical method of pair production for
  arbitrary spins and photon polarization},}\ }\href {\doibase
  10.1103/PhysRevD.101.076017} {\bibfield  {journal} {\bibinfo  {journal}
  {Phys. Rev. D}\ }\textbf {\bibinfo {volume} {101}},\ \bibinfo {pages}
  {076017} (\bibinfo {year} {2020})}\BibitemShut {NoStop}%
\bibitem [{\citenamefont {Wan}\ \emph {et~al.}(2020)\citenamefont {Wan},
  \citenamefont {Wang}, \citenamefont {Guo}, \citenamefont {Chen},
  \citenamefont {Shaisultanov}, \citenamefont {Xu}, \citenamefont
  {Hatsagortsyan}, \citenamefont {Keitel},\ and\ \citenamefont
  {Li}}]{Wan_2020}%
  \BibitemOpen
  \bibfield  {author} {\bibinfo {author} {\bibfnamefont {Feng}\ \bibnamefont
  {Wan}}, \bibinfo {author} {\bibfnamefont {Yu}~\bibnamefont {Wang}}, \bibinfo
  {author} {\bibfnamefont {Ren-Tong}\ \bibnamefont {Guo}}, \bibinfo {author}
  {\bibfnamefont {Yue-Yue}\ \bibnamefont {Chen}}, \bibinfo {author}
  {\bibfnamefont {Rashid}\ \bibnamefont {Shaisultanov}}, \bibinfo {author}
  {\bibfnamefont {Zhong-Feng}\ \bibnamefont {Xu}}, \bibinfo {author}
  {\bibfnamefont {Karen~Z.}\ \bibnamefont {Hatsagortsyan}}, \bibinfo {author}
  {\bibfnamefont {Christoph~H.}\ \bibnamefont {Keitel}}, \ and\ \bibinfo
  {author} {\bibfnamefont {Jian-Xing}\ \bibnamefont {Li}},\ }\bibfield  {title}
  {\enquote {\bibinfo {title} {{High-energy $\ensuremath{\gamma}$-photon
  polarization in nonlinear Breit-Wheeler pair production and
  $\ensuremath{\gamma}$ polarimetry}},}\ }\href {\doibase
  10.1103/PhysRevResearch.2.032049} {\bibfield  {journal} {\bibinfo  {journal}
  {Phys. Rev. Research}\ }\textbf {\bibinfo {volume} {2}},\ \bibinfo {pages}
  {032049} (\bibinfo {year} {2020})}\BibitemShut {NoStop}%
\bibitem [{\citenamefont {Del~Sorbo}\ \emph {et~al.}(2017)\citenamefont
  {Del~Sorbo}, \citenamefont {Seipt}, \citenamefont {Blackburn}, \citenamefont
  {Thomas}, \citenamefont {Murphy}, \citenamefont {Kirk},\ and\ \citenamefont
  {Ridgers}}]{Sorbo_2017}%
  \BibitemOpen
  \bibfield  {author} {\bibinfo {author} {\bibfnamefont {D.}~\bibnamefont
  {Del~Sorbo}}, \bibinfo {author} {\bibfnamefont {D.}~\bibnamefont {Seipt}},
  \bibinfo {author} {\bibfnamefont {T.~G.}\ \bibnamefont {Blackburn}}, \bibinfo
  {author} {\bibfnamefont {A.~G.~R.}\ \bibnamefont {Thomas}}, \bibinfo {author}
  {\bibfnamefont {C.~D.}\ \bibnamefont {Murphy}}, \bibinfo {author}
  {\bibfnamefont {J.~G.}\ \bibnamefont {Kirk}}, \ and\ \bibinfo {author}
  {\bibfnamefont {C.~P.}\ \bibnamefont {Ridgers}},\ }\bibfield  {title}
  {\enquote {\bibinfo {title} {Spin polarization of electrons by ultraintense
  lasers},}\ }\href {\doibase 10.1103/PhysRevA.96.043407} {\bibfield  {journal}
  {\bibinfo  {journal} {Phys. Rev. A}\ }\textbf {\bibinfo {volume} {96}},\
  \bibinfo {pages} {043407} (\bibinfo {year} {2017})}\BibitemShut {NoStop}%
\bibitem [{\citenamefont {Del~Sorbo}\ \emph {et~al.}(2018)\citenamefont
  {Del~Sorbo}, \citenamefont {Seipt}, \citenamefont {Thomas},\ and\
  \citenamefont {Ridgers}}]{Sorbo_2018}%
  \BibitemOpen
  \bibfield  {author} {\bibinfo {author} {\bibfnamefont {D.}~\bibnamefont
  {Del~Sorbo}}, \bibinfo {author} {\bibfnamefont {D.}~\bibnamefont {Seipt}},
  \bibinfo {author} {\bibfnamefont {A.~G.~R.}\ \bibnamefont {Thomas}}, \ and\
  \bibinfo {author} {\bibfnamefont {C.~P.}\ \bibnamefont {Ridgers}},\
  }\bibfield  {title} {\enquote {\bibinfo {title} {Electron spin polarization
  in realistic trajectories around the magnetic node of two
  counter-propagating, circularly polarized, ultra-intense lasers},}\
  }\href@noop {} {\bibfield  {journal} {\bibinfo  {journal} {Plasma Phys.
  Control. Fusion}\ }\textbf {\bibinfo {volume} {60}},\ \bibinfo {pages}
  {064003} (\bibinfo {year} {2018})}\BibitemShut {NoStop}%
\bibitem [{\citenamefont {Seipt}\ \emph {et~al.}(2018)\citenamefont {Seipt},
  \citenamefont {Del~Sorbo}, \citenamefont {Ridgers},\ and\ \citenamefont
  {Thomas}}]{Seipt_2018}%
  \BibitemOpen
  \bibfield  {author} {\bibinfo {author} {\bibfnamefont {D.}~\bibnamefont
  {Seipt}}, \bibinfo {author} {\bibfnamefont {D.}~\bibnamefont {Del~Sorbo}},
  \bibinfo {author} {\bibfnamefont {C.~P.}\ \bibnamefont {Ridgers}}, \ and\
  \bibinfo {author} {\bibfnamefont {A.~G.~R.}\ \bibnamefont {Thomas}},\
  }\bibfield  {title} {\enquote {\bibinfo {title} {Theory of radiative electron
  polarization in strong laser fields},}\ }\href@noop {} {\bibfield  {journal}
  {\bibinfo  {journal} {Phys. Rev. A}\ }\textbf {\bibinfo {volume} {98}},\
  \bibinfo {pages} {023417} (\bibinfo {year} {2018})}\BibitemShut {NoStop}%
\bibitem [{\citenamefont {Li}\ \emph {et~al.}(2019)\citenamefont {Li},
  \citenamefont {Shaisultanov}, \citenamefont {Hatsagortsyan}, \citenamefont
  {Wan}, \citenamefont {Keitel},\ and\ \citenamefont {Li}}]{li2019prl}%
  \BibitemOpen
  \bibfield  {author} {\bibinfo {author} {\bibfnamefont {Yan-Fei}\ \bibnamefont
  {Li}}, \bibinfo {author} {\bibfnamefont {Rashid}\ \bibnamefont
  {Shaisultanov}}, \bibinfo {author} {\bibfnamefont {Karen~Z.}\ \bibnamefont
  {Hatsagortsyan}}, \bibinfo {author} {\bibfnamefont {Feng}\ \bibnamefont
  {Wan}}, \bibinfo {author} {\bibfnamefont {Christoph~H.}\ \bibnamefont
  {Keitel}}, \ and\ \bibinfo {author} {\bibfnamefont {Jian-Xing}\ \bibnamefont
  {Li}},\ }\bibfield  {title} {\enquote {\bibinfo {title} {Ultrarelativistic
  electron-beam polarization in single-shot interaction with an ultraintense
  laser pulse},}\ }\href {\doibase 10.1103/PhysRevLett.122.154801} {\bibfield
  {journal} {\bibinfo  {journal} {Phys. Rev. Lett.}\ }\textbf {\bibinfo
  {volume} {122}},\ \bibinfo {pages} {154801} (\bibinfo {year}
  {2019})}\BibitemShut {NoStop}%
\bibitem [{\citenamefont {Seipt}\ \emph {et~al.}(2019)\citenamefont {Seipt},
  \citenamefont {Del~Sorbo}, \citenamefont {Ridgers},\ and\ \citenamefont
  {Thomas}}]{Seipt_2019}%
  \BibitemOpen
  \bibfield  {author} {\bibinfo {author} {\bibfnamefont {Daniel}\ \bibnamefont
  {Seipt}}, \bibinfo {author} {\bibfnamefont {Dario}\ \bibnamefont
  {Del~Sorbo}}, \bibinfo {author} {\bibfnamefont {Christopher~P.}\ \bibnamefont
  {Ridgers}}, \ and\ \bibinfo {author} {\bibfnamefont {Alec G.~R.}\
  \bibnamefont {Thomas}},\ }\bibfield  {title} {\enquote {\bibinfo {title}
  {Ultrafast polarization of an electron beam in an intense bichromatic laser
  field},}\ }\href {\doibase 10.1103/PhysRevA.100.061402} {\bibfield  {journal}
  {\bibinfo  {journal} {Phys. Rev. A}\ }\textbf {\bibinfo {volume} {100}},\
  \bibinfo {pages} {061402(R)} (\bibinfo {year} {2019})}\BibitemShut {NoStop}%
\bibitem [{\citenamefont {Song}\ \emph {et~al.}(2019)\citenamefont {Song},
  \citenamefont {Wang}, \citenamefont {Li}, \citenamefont {Li},\ and\
  \citenamefont {Li}}]{Song_2019}%
  \BibitemOpen
  \bibfield  {author} {\bibinfo {author} {\bibfnamefont {Huai-Hang}\
  \bibnamefont {Song}}, \bibinfo {author} {\bibfnamefont {Wei-Min}\
  \bibnamefont {Wang}}, \bibinfo {author} {\bibfnamefont {Jian-Xing}\
  \bibnamefont {Li}}, \bibinfo {author} {\bibfnamefont {Yan-Fei}\ \bibnamefont
  {Li}}, \ and\ \bibinfo {author} {\bibfnamefont {Yu-Tong}\ \bibnamefont
  {Li}},\ }\bibfield  {title} {\enquote {\bibinfo {title} {Spin-polarization
  effects of an ultrarelativistic electron beam in an ultraintense two-color
  laser pulse},}\ }\href {\doibase 10.1103/PhysRevA.100.033407} {\bibfield
  {journal} {\bibinfo  {journal} {Phys. Rev. A}\ }\textbf {\bibinfo {volume}
  {100}},\ \bibinfo {pages} {033407} (\bibinfo {year} {2019})}\BibitemShut
  {NoStop}%
\bibitem [{\citenamefont {Chen}\ \emph {et~al.}(2019)\citenamefont {Chen},
  \citenamefont {He}, \citenamefont {Shaisultanov}, \citenamefont
  {Hatsagortsyan},\ and\ \citenamefont {Keitel}}]{Chen_2019}%
  \BibitemOpen
  \bibfield  {author} {\bibinfo {author} {\bibfnamefont {Yue-Yue}\ \bibnamefont
  {Chen}}, \bibinfo {author} {\bibfnamefont {Pei-Lun}\ \bibnamefont {He}},
  \bibinfo {author} {\bibfnamefont {Rashid}\ \bibnamefont {Shaisultanov}},
  \bibinfo {author} {\bibfnamefont {Karen~Z.}\ \bibnamefont {Hatsagortsyan}}, \
  and\ \bibinfo {author} {\bibfnamefont {Christoph~H.}\ \bibnamefont
  {Keitel}},\ }\bibfield  {title} {\enquote {\bibinfo {title} {Polarized
  positron beams via intense two-color laser pulses},}\ }\href {\doibase
  10.1103/PhysRevLett.123.174801} {\bibfield  {journal} {\bibinfo  {journal}
  {Phys. Rev. Lett.}\ }\textbf {\bibinfo {volume} {123}},\ \bibinfo {pages}
  {174801} (\bibinfo {year} {2019})}\BibitemShut {NoStop}%
\bibitem [{\citenamefont {Kotkin}\ \emph {et~al.}(2003)\citenamefont {Kotkin},
  \citenamefont {Serbo},\ and\ \citenamefont {Telnov}}]{Kotkin2003prstab}%
  \BibitemOpen
  \bibfield  {author} {\bibinfo {author} {\bibfnamefont {G.~L.}\ \bibnamefont
  {Kotkin}}, \bibinfo {author} {\bibfnamefont {V.~G.}\ \bibnamefont {Serbo}}, \
  and\ \bibinfo {author} {\bibfnamefont {V.~I.}\ \bibnamefont {Telnov}},\
  }\bibfield  {title} {\enquote {\bibinfo {title} {Electron (positron) beam
  polarization by compton scattering on circularly polarized laser photons},}\
  }\href {\doibase 10.1103/PhysRevSTAB.6.011001} {\bibfield  {journal}
  {\bibinfo  {journal} {Phys. Rev. ST Accel. Beams}\ }\textbf {\bibinfo
  {volume} {6}},\ \bibinfo {pages} {011001} (\bibinfo {year}
  {2003})}\BibitemShut {NoStop}%
\bibitem [{\citenamefont {Ivanov}\ \emph {et~al.}(2004)\citenamefont {Ivanov},
  \citenamefont {Kotkin},\ and\ \citenamefont {Serbo}}]{Ivanov_2004}%
  \BibitemOpen
  \bibfield  {author} {\bibinfo {author} {\bibfnamefont {D.~Yu.}\ \bibnamefont
  {Ivanov}}, \bibinfo {author} {\bibfnamefont {G.~L.}\ \bibnamefont {Kotkin}},
  \ and\ \bibinfo {author} {\bibfnamefont {V.~G.}\ \bibnamefont {Serbo}},\
  }\bibfield  {title} {\enquote {\bibinfo {title} {Complete description of
  polarization effects in emission of a photon by an electron in the field of a
  strong laser wave},}\ }\href {\doibase 10.1140/epjc/s2004-01861-x} {\bibfield
   {journal} {\bibinfo  {journal} {Eur. Phys. J. C}\ }\textbf {\bibinfo
  {volume} {36}},\ \bibinfo {pages} {127--145} (\bibinfo {year}
  {2004})}\BibitemShut {NoStop}%
\bibitem [{\citenamefont {Karlovets}(2011)}]{Karlovets_2011}%
  \BibitemOpen
  \bibfield  {author} {\bibinfo {author} {\bibfnamefont {Dmitry~V.}\
  \bibnamefont {Karlovets}},\ }\bibfield  {title} {\enquote {\bibinfo {title}
  {Radiative polarization of electrons in a strong laser wave},}\ }\href
  {\doibase 10.1103/PhysRevA.84.062116} {\bibfield  {journal} {\bibinfo
  {journal} {Phys. Rev. A}\ }\textbf {\bibinfo {volume} {84}},\ \bibinfo
  {pages} {062116} (\bibinfo {year} {2011})}\BibitemShut {NoStop}%
\bibitem [{\citenamefont {Li}\ \emph {et~al.}(2020{\natexlab{a}})\citenamefont
  {Li}, \citenamefont {Chen}, \citenamefont {Wang},\ and\ \citenamefont
  {Hu}}]{Li_2020_2}%
  \BibitemOpen
  \bibfield  {author} {\bibinfo {author} {\bibfnamefont {Yan-Fei}\ \bibnamefont
  {Li}}, \bibinfo {author} {\bibfnamefont {Yue-Yue}\ \bibnamefont {Chen}},
  \bibinfo {author} {\bibfnamefont {Wei-Min}\ \bibnamefont {Wang}}, \ and\
  \bibinfo {author} {\bibfnamefont {Hua-Si}\ \bibnamefont {Hu}},\ }\bibfield
  {title} {\enquote {\bibinfo {title} {Production of highly polarized positron
  beams via helicity transfer from polarized electrons in a strong laser
  field},}\ }\href {\doibase 10.1103/PhysRevLett.125.044802} {\bibfield
  {journal} {\bibinfo  {journal} {Phys. Rev. Lett.}\ }\textbf {\bibinfo
  {volume} {125}},\ \bibinfo {pages} {044802} (\bibinfo {year}
  {2020}{\natexlab{a}})}\BibitemShut {NoStop}%
\bibitem [{\citenamefont {Pierce}\ \emph {et~al.}(1980)\citenamefont {Pierce},
  \citenamefont {Celotta}, \citenamefont {Wang}, \citenamefont {Unertl},
  \citenamefont {Galejs}, \citenamefont {Kuyatt},\ and\ \citenamefont
  {Mielczarek}}]{Pierce1980}%
  \BibitemOpen
  \bibfield  {author} {\bibinfo {author} {\bibfnamefont {D.~T.}\ \bibnamefont
  {Pierce}}, \bibinfo {author} {\bibfnamefont {R.~J.}\ \bibnamefont {Celotta}},
  \bibinfo {author} {\bibfnamefont {G.-C.}\ \bibnamefont {Wang}}, \bibinfo
  {author} {\bibfnamefont {W.~N.}\ \bibnamefont {Unertl}}, \bibinfo {author}
  {\bibfnamefont {A.}~\bibnamefont {Galejs}}, \bibinfo {author} {\bibfnamefont
  {C.~E.}\ \bibnamefont {Kuyatt}}, \ and\ \bibinfo {author} {\bibfnamefont
  {S.~R.}\ \bibnamefont {Mielczarek}},\ }\bibfield  {title} {\enquote {\bibinfo
  {title} {{The GaAs spin polarized electron source}},}\ }\href {\doibase
  10.1063/1.1136250} {\bibfield  {journal} {\bibinfo  {journal} {Rev. Sci.
  Instrum.}\ }\textbf {\bibinfo {volume} {51}},\ \bibinfo {pages} {478--499}
  (\bibinfo {year} {1980})}\BibitemShut {NoStop}%
\bibitem [{\citenamefont {Kuwahara}\ \emph {et~al.}(2012)\citenamefont
  {Kuwahara}, \citenamefont {Kusunoki}, \citenamefont {Jin}, \citenamefont
  {Nakanishi}, \citenamefont {Takeda}, \citenamefont {Saitoh}, \citenamefont
  {Ujihara}, \citenamefont {Asano},\ and\ \citenamefont
  {Tanaka}}]{Kuwahara2012}%
  \BibitemOpen
  \bibfield  {author} {\bibinfo {author} {\bibfnamefont {M.}~\bibnamefont
  {Kuwahara}}, \bibinfo {author} {\bibfnamefont {S.}~\bibnamefont {Kusunoki}},
  \bibinfo {author} {\bibfnamefont {X.~G.}\ \bibnamefont {Jin}}, \bibinfo
  {author} {\bibfnamefont {T.}~\bibnamefont {Nakanishi}}, \bibinfo {author}
  {\bibfnamefont {Y.}~\bibnamefont {Takeda}}, \bibinfo {author} {\bibfnamefont
  {K.}~\bibnamefont {Saitoh}}, \bibinfo {author} {\bibfnamefont
  {T.}~\bibnamefont {Ujihara}}, \bibinfo {author} {\bibfnamefont
  {H.}~\bibnamefont {Asano}}, \ and\ \bibinfo {author} {\bibfnamefont
  {N.}~\bibnamefont {Tanaka}},\ }\bibfield  {title} {\enquote {\bibinfo {title}
  {{30-kV spin-polarized transmission electron microscope with GaAs-GaAsP
  strained superlattice photocathode}},}\ }\href {\doibase 10.1063/1.4737177}
  {\bibfield  {journal} {\bibinfo  {journal} {Appl. Phys. Lett.}\ }\textbf
  {\bibinfo {volume} {101}},\ \bibinfo {pages} {033102} (\bibinfo {year}
  {2012})}\BibitemShut {NoStop}%
\bibitem [{\citenamefont {Zitzewitz}\ \emph {et~al.}(1979)\citenamefont
  {Zitzewitz}, \citenamefont {Van~House}, \citenamefont {Rich},\ and\
  \citenamefont {Gidley}}]{Zitzewitz1979}%
  \BibitemOpen
  \bibfield  {author} {\bibinfo {author} {\bibfnamefont {P.~W.}\ \bibnamefont
  {Zitzewitz}}, \bibinfo {author} {\bibfnamefont {J.~C.}\ \bibnamefont
  {Van~House}}, \bibinfo {author} {\bibfnamefont {A.}~\bibnamefont {Rich}}, \
  and\ \bibinfo {author} {\bibfnamefont {D.~W.}\ \bibnamefont {Gidley}},\
  }\bibfield  {title} {\enquote {\bibinfo {title} {Spin polarization of
  low-energy positron beams},}\ }\href {\doibase 10.1103/PhysRevLett.43.1281}
  {\bibfield  {journal} {\bibinfo  {journal} {Phys. Rev. Lett.}\ }\textbf
  {\bibinfo {volume} {43}},\ \bibinfo {pages} {1281--1284} (\bibinfo {year}
  {1979})}\BibitemShut {NoStop}%
\bibitem [{\citenamefont {Barth}\ and\ \citenamefont
  {Smirnova}(2013)}]{Barth2013}%
  \BibitemOpen
  \bibfield  {author} {\bibinfo {author} {\bibfnamefont {Ingo}\ \bibnamefont
  {Barth}}\ and\ \bibinfo {author} {\bibfnamefont {Olga}\ \bibnamefont
  {Smirnova}},\ }\bibfield  {title} {\enquote {\bibinfo {title} {Spin-polarized
  electrons produced by strong-field ionization},}\ }\href {\doibase
  10.1103/PhysRevA.88.013401} {\bibfield  {journal} {\bibinfo  {journal} {Phys.
  Rev. A}\ }\textbf {\bibinfo {volume} {88}},\ \bibinfo {pages} {013401}
  (\bibinfo {year} {2013})}\BibitemShut {NoStop}%
\bibitem [{\citenamefont {Rakitzis}\ \emph {et~al.}(2003)\citenamefont
  {Rakitzis}, \citenamefont {Samartzis}, \citenamefont {Toomes}, \citenamefont
  {Kitsopoulos}, \citenamefont {Brown}, \citenamefont {Balint-Kurti},
  \citenamefont {Vasyutinskii},\ and\ \citenamefont {Beswick}}]{Rakitzis1936}%
  \BibitemOpen
  \bibfield  {author} {\bibinfo {author} {\bibfnamefont {T.~P.}\ \bibnamefont
  {Rakitzis}}, \bibinfo {author} {\bibfnamefont {P.~C.}\ \bibnamefont
  {Samartzis}}, \bibinfo {author} {\bibfnamefont {R.~L.}\ \bibnamefont
  {Toomes}}, \bibinfo {author} {\bibfnamefont {T.~N.}\ \bibnamefont
  {Kitsopoulos}}, \bibinfo {author} {\bibfnamefont {Alex}\ \bibnamefont
  {Brown}}, \bibinfo {author} {\bibfnamefont {G.~G.}\ \bibnamefont
  {Balint-Kurti}}, \bibinfo {author} {\bibfnamefont {O.~S.}\ \bibnamefont
  {Vasyutinskii}}, \ and\ \bibinfo {author} {\bibfnamefont {J.~A.}\
  \bibnamefont {Beswick}},\ }\bibfield  {title} {\enquote {\bibinfo {title}
  {{Spin-Polarized Hydrogen Atoms from Molecular Photodissociation}},}\ }\href
  {\doibase 10.1126/science.1084809} {\bibfield  {journal} {\bibinfo  {journal}
  {Science}\ }\textbf {\bibinfo {volume} {300}},\ \bibinfo {pages} {1936--1938}
  (\bibinfo {year} {2003})}\BibitemShut {NoStop}%
\bibitem [{\citenamefont {Sofikitis}\ \emph {et~al.}(2017)\citenamefont
  {Sofikitis}, \citenamefont {Glodic}, \citenamefont {Koumarianou},
  \citenamefont {Jiang}, \citenamefont {Bougas}, \citenamefont {Samartzis},
  \citenamefont {Andreev},\ and\ \citenamefont {Rakitzis}}]{sofikitis2017}%
  \BibitemOpen
  \bibfield  {author} {\bibinfo {author} {\bibfnamefont {Dimitris}\
  \bibnamefont {Sofikitis}}, \bibinfo {author} {\bibfnamefont {Pavle}\
  \bibnamefont {Glodic}}, \bibinfo {author} {\bibfnamefont {Greta}\
  \bibnamefont {Koumarianou}}, \bibinfo {author} {\bibfnamefont {Hongyan}\
  \bibnamefont {Jiang}}, \bibinfo {author} {\bibfnamefont {Lykourgos}\
  \bibnamefont {Bougas}}, \bibinfo {author} {\bibfnamefont {Peter~C.}\
  \bibnamefont {Samartzis}}, \bibinfo {author} {\bibfnamefont {Alexander}\
  \bibnamefont {Andreev}}, \ and\ \bibinfo {author} {\bibfnamefont {T.~Peter}\
  \bibnamefont {Rakitzis}},\ }\bibfield  {title} {\enquote {\bibinfo {title}
  {Highly nuclear-spin-polarized deuterium atoms from the uv photodissociation
  of deuterium iodide},}\ }\href {\doibase 10.1103/PhysRevLett.118.233401}
  {\bibfield  {journal} {\bibinfo  {journal} {Phys. Rev. Lett.}\ }\textbf
  {\bibinfo {volume} {118}},\ \bibinfo {pages} {233401} (\bibinfo {year}
  {2017})}\BibitemShut {NoStop}%
\bibitem [{\citenamefont {Sofikitis}\ \emph {et~al.}(2018)\citenamefont
  {Sofikitis}, \citenamefont {Kannis}, \citenamefont {Boulogiannis},\ and\
  \citenamefont {Rakitzis}}]{sofikitis2018}%
  \BibitemOpen
  \bibfield  {author} {\bibinfo {author} {\bibfnamefont {Dimitris}\
  \bibnamefont {Sofikitis}}, \bibinfo {author} {\bibfnamefont
  {Chrysovalantis~S.}\ \bibnamefont {Kannis}}, \bibinfo {author} {\bibfnamefont
  {Gregoris~K.}\ \bibnamefont {Boulogiannis}}, \ and\ \bibinfo {author}
  {\bibfnamefont {T.~Peter}\ \bibnamefont {Rakitzis}},\ }\bibfield  {title}
  {\enquote {\bibinfo {title} {Ultrahigh-density spin-polarized h and d
  observed via magnetization quantum beats},}\ }\href {\doibase
  10.1103/PhysRevLett.121.083001} {\bibfield  {journal} {\bibinfo  {journal}
  {Phys. Rev. Lett.}\ }\textbf {\bibinfo {volume} {121}},\ \bibinfo {pages}
  {083001} (\bibinfo {year} {2018})}\BibitemShut {NoStop}%
\bibitem [{\citenamefont {Wen}\ \emph {et~al.}(2019)\citenamefont {Wen},
  \citenamefont {Tamburini},\ and\ \citenamefont {Keitel}}]{Wen_2019}%
  \BibitemOpen
  \bibfield  {author} {\bibinfo {author} {\bibfnamefont {Meng}\ \bibnamefont
  {Wen}}, \bibinfo {author} {\bibfnamefont {Matteo}\ \bibnamefont {Tamburini}},
  \ and\ \bibinfo {author} {\bibfnamefont {Christoph~H.}\ \bibnamefont
  {Keitel}},\ }\bibfield  {title} {\enquote {\bibinfo {title} {Polarized
  laser-wakefield-accelerated kiloampere electron beams},}\ }\href {\doibase
  10.1103/PhysRevLett.122.214801} {\bibfield  {journal} {\bibinfo  {journal}
  {Phys. Rev. Lett.}\ }\textbf {\bibinfo {volume} {122}},\ \bibinfo {pages}
  {214801} (\bibinfo {year} {2019})}\BibitemShut {NoStop}%
\bibitem [{\citenamefont {Wu}\ \emph {et~al.}(2019{\natexlab{a}})\citenamefont
  {Wu}, \citenamefont {Ji}, \citenamefont {Geng}, \citenamefont {Yu},
  \citenamefont {Wang}, \citenamefont {Feng}, \citenamefont {Guo},
  \citenamefont {Wang}, \citenamefont {Qin}, \citenamefont {Yan}, \citenamefont
  {Zhang}, \citenamefont {Thomas}, \citenamefont {Hützen}, \citenamefont
  {Büscher}, \citenamefont {Rakitzis}, \citenamefont {Pukhov}, \citenamefont
  {Shen},\ and\ \citenamefont {Li}}]{Yitong2019b}%
  \BibitemOpen
  \bibfield  {author} {\bibinfo {author} {\bibfnamefont {Yitong}\ \bibnamefont
  {Wu}}, \bibinfo {author} {\bibfnamefont {Liangliang}\ \bibnamefont {Ji}},
  \bibinfo {author} {\bibfnamefont {Xuesong}\ \bibnamefont {Geng}}, \bibinfo
  {author} {\bibfnamefont {Qin}\ \bibnamefont {Yu}}, \bibinfo {author}
  {\bibfnamefont {Nengwen}\ \bibnamefont {Wang}}, \bibinfo {author}
  {\bibfnamefont {Bo}~\bibnamefont {Feng}}, \bibinfo {author} {\bibfnamefont
  {Zhao}\ \bibnamefont {Guo}}, \bibinfo {author} {\bibfnamefont {Weiqing}\
  \bibnamefont {Wang}}, \bibinfo {author} {\bibfnamefont {Chengyu}\
  \bibnamefont {Qin}}, \bibinfo {author} {\bibfnamefont {Xue}\ \bibnamefont
  {Yan}}, \bibinfo {author} {\bibfnamefont {Lingang}\ \bibnamefont {Zhang}},
  \bibinfo {author} {\bibfnamefont {Johannes}\ \bibnamefont {Thomas}}, \bibinfo
  {author} {\bibfnamefont {Anna}\ \bibnamefont {Hützen}}, \bibinfo {author}
  {\bibfnamefont {Markus}\ \bibnamefont {Büscher}}, \bibinfo {author}
  {\bibfnamefont {T~Peter}\ \bibnamefont {Rakitzis}}, \bibinfo {author}
  {\bibfnamefont {Alexander}\ \bibnamefont {Pukhov}}, \bibinfo {author}
  {\bibfnamefont {Baifei}\ \bibnamefont {Shen}}, \ and\ \bibinfo {author}
  {\bibfnamefont {Ruxin}\ \bibnamefont {Li}},\ }\bibfield  {title} {\enquote
  {\bibinfo {title} {Polarized electron-beam acceleration driven by vortex
  laser pulses},}\ }\href {\doibase 10.1088/1367-2630/ab2fd7} {\bibfield
  {journal} {\bibinfo  {journal} {New Journal of Physics}\ }\textbf {\bibinfo
  {volume} {21}},\ \bibinfo {pages} {073052} (\bibinfo {year}
  {2019}{\natexlab{a}})}\BibitemShut {NoStop}%
\bibitem [{\citenamefont {Wu}\ \emph {et~al.}(2019{\natexlab{b}})\citenamefont
  {Wu}, \citenamefont {Ji}, \citenamefont {Geng}, \citenamefont {Yu},
  \citenamefont {Wang}, \citenamefont {Feng}, \citenamefont {Guo},
  \citenamefont {Wang}, \citenamefont {Qin}, \citenamefont {Yan}, \citenamefont
  {Zhang}, \citenamefont {Thomas}, \citenamefont {H\"utzen}, \citenamefont
  {Pukhov}, \citenamefont {B\"uscher}, \citenamefont {Shen},\ and\
  \citenamefont {Li}}]{Yitong2019a}%
  \BibitemOpen
  \bibfield  {author} {\bibinfo {author} {\bibfnamefont {Yitong}\ \bibnamefont
  {Wu}}, \bibinfo {author} {\bibfnamefont {Liangliang}\ \bibnamefont {Ji}},
  \bibinfo {author} {\bibfnamefont {Xuesong}\ \bibnamefont {Geng}}, \bibinfo
  {author} {\bibfnamefont {Qin}\ \bibnamefont {Yu}}, \bibinfo {author}
  {\bibfnamefont {Nengwen}\ \bibnamefont {Wang}}, \bibinfo {author}
  {\bibfnamefont {Bo}~\bibnamefont {Feng}}, \bibinfo {author} {\bibfnamefont
  {Zhao}\ \bibnamefont {Guo}}, \bibinfo {author} {\bibfnamefont {Weiqing}\
  \bibnamefont {Wang}}, \bibinfo {author} {\bibfnamefont {Chengyu}\
  \bibnamefont {Qin}}, \bibinfo {author} {\bibfnamefont {Xue}\ \bibnamefont
  {Yan}}, \bibinfo {author} {\bibfnamefont {Lingang}\ \bibnamefont {Zhang}},
  \bibinfo {author} {\bibfnamefont {Johannes}\ \bibnamefont {Thomas}}, \bibinfo
  {author} {\bibfnamefont {Anna}\ \bibnamefont {H\"utzen}}, \bibinfo {author}
  {\bibfnamefont {Alexander}\ \bibnamefont {Pukhov}}, \bibinfo {author}
  {\bibfnamefont {Markus}\ \bibnamefont {B\"uscher}}, \bibinfo {author}
  {\bibfnamefont {Baifei}\ \bibnamefont {Shen}}, \ and\ \bibinfo {author}
  {\bibfnamefont {Ruxin}\ \bibnamefont {Li}},\ }\bibfield  {title} {\enquote
  {\bibinfo {title} {Polarized electron acceleration in beam-driven plasma
  wakefield based on density down-ramp injection},}\ }\href {\doibase
  10.1103/PhysRevE.100.043202} {\bibfield  {journal} {\bibinfo  {journal}
  {Phys. Rev. E}\ }\textbf {\bibinfo {volume} {100}},\ \bibinfo {pages}
  {043202} (\bibinfo {year} {2019}{\natexlab{b}})}\BibitemShut {NoStop}%
\bibitem [{\citenamefont {Nie}\ \emph {et~al.}(2021)\citenamefont {Nie},
  \citenamefont {Li}, \citenamefont {Morales}, \citenamefont {Patchkovskii},
  \citenamefont {Smirnova}, \citenamefont {An}, \citenamefont {Nambu},
  \citenamefont {Matteo}, \citenamefont {Marsh}, \citenamefont {Tsung},
  \citenamefont {Mori},\ and\ \citenamefont {Joshi}}]{Nie2021}%
  \BibitemOpen
  \bibfield  {author} {\bibinfo {author} {\bibfnamefont {Zan}\ \bibnamefont
  {Nie}}, \bibinfo {author} {\bibfnamefont {Fei}\ \bibnamefont {Li}}, \bibinfo
  {author} {\bibfnamefont {Felipe}\ \bibnamefont {Morales}}, \bibinfo {author}
  {\bibfnamefont {Serguei}\ \bibnamefont {Patchkovskii}}, \bibinfo {author}
  {\bibfnamefont {Olga}\ \bibnamefont {Smirnova}}, \bibinfo {author}
  {\bibfnamefont {Weiming}\ \bibnamefont {An}}, \bibinfo {author}
  {\bibfnamefont {Noa}\ \bibnamefont {Nambu}}, \bibinfo {author} {\bibfnamefont
  {Daniel}\ \bibnamefont {Matteo}}, \bibinfo {author} {\bibfnamefont
  {Kenneth~A.}\ \bibnamefont {Marsh}}, \bibinfo {author} {\bibfnamefont
  {Frank}\ \bibnamefont {Tsung}}, \bibinfo {author} {\bibfnamefont {Warren~B.}\
  \bibnamefont {Mori}}, \ and\ \bibinfo {author} {\bibfnamefont {Chan}\
  \bibnamefont {Joshi}},\ }\bibfield  {title} {\enquote {\bibinfo {title} {{ In
  Situ Generation of High-Energy Spin-Polarized Electrons in a Beam-Driven
  Plasma Wakefield Accelerator }},}\ }\href {\doibase
  10.1103/physrevlett.126.054801} {\bibfield  {journal} {\bibinfo  {journal}
  {Physical Review Letters}\ }\textbf {\bibinfo {volume} {126}},\ \bibinfo
  {pages} {54801} (\bibinfo {year} {2021})},\ \Eprint
  {http://arxiv.org/abs/2101.10378} {2101.10378} \BibitemShut {NoStop}%
\bibitem [{\citenamefont {Baier}\ \emph {et~al.}(1998)\citenamefont {Baier},
  \citenamefont {Katkov},\ and\ \citenamefont {Strakhovenko}}]{Baier1998}%
  \BibitemOpen
  \bibfield  {author} {\bibinfo {author} {\bibfnamefont {V.~N.}\ \bibnamefont
  {Baier}}, \bibinfo {author} {\bibfnamefont {V.~M.}\ \bibnamefont {Katkov}}, \
  and\ \bibinfo {author} {\bibfnamefont {V.~M.}\ \bibnamefont {Strakhovenko}},\
  }\href@noop {} {\emph {\bibinfo {title} {Electromagnetic Processes at High
  Energies in Oriented Single Crystals}}}\ (\bibinfo  {publisher} {World
  Scientific},\ \bibinfo {address} {Singapore},\ \bibinfo {year}
  {1998})\BibitemShut {NoStop}%
\bibitem [{\citenamefont {McMaster}(1961)}]{McMaster_1961}%
  \BibitemOpen
  \bibfield  {author} {\bibinfo {author} {\bibfnamefont {William~H.}\
  \bibnamefont {McMaster}},\ }\bibfield  {title} {\enquote {\bibinfo {title}
  {Matrix representation of polarization},}\ }\href {\doibase
  10.1103/RevModPhys.33.8} {\bibfield  {journal} {\bibinfo  {journal} {Rev.
  Mod. Phys.}\ }\textbf {\bibinfo {volume} {33}},\ \bibinfo {pages} {8--28}
  (\bibinfo {year} {1961})}\BibitemShut {NoStop}%
\bibitem [{\citenamefont {Boyarkin}(2011)}]{Boyarkin2011}%
  \BibitemOpen
  \bibfield  {author} {\bibinfo {author} {\bibfnamefont {O.M.}\ \bibnamefont
  {Boyarkin}},\ }\href@noop {} {\emph {\bibinfo {title} {Advanced particle
  physics volume I: Particles, fields, and quantum electrodynamics}}}\
  (\bibinfo {year} {2011})\BibitemShut {NoStop}%
\bibitem [{\citenamefont {Baier}\ \emph {et~al.}(1973)\citenamefont {Baier},
  \citenamefont {Katkov},\ and\ \citenamefont {Fadin}}]{Baier_1973}%
  \BibitemOpen
  \bibfield  {author} {\bibinfo {author} {\bibfnamefont {V.~N.}\ \bibnamefont
  {Baier}}, \bibinfo {author} {\bibfnamefont {V.~M.}\ \bibnamefont {Katkov}}, \
  and\ \bibinfo {author} {\bibfnamefont {V.~S.}\ \bibnamefont {Fadin}},\
  }\href@noop {} {\emph {\bibinfo {title} {Radiation from relativistic
  electrons}}}\ (\bibinfo  {publisher} {Atomizdat, Moscow},\ \bibinfo {year}
  {1973})\BibitemShut {NoStop}%
\bibitem [{\citenamefont {Di~Piazza}\ \emph {et~al.}(2018)\citenamefont
  {Di~Piazza}, \citenamefont {Tamburini}, \citenamefont {Meuren},\ and\
  \citenamefont {Keitel}}]{Piazza_2018}%
  \BibitemOpen
  \bibfield  {author} {\bibinfo {author} {\bibfnamefont {A.}~\bibnamefont
  {Di~Piazza}}, \bibinfo {author} {\bibfnamefont {M.}~\bibnamefont
  {Tamburini}}, \bibinfo {author} {\bibfnamefont {S.}~\bibnamefont {Meuren}}, \
  and\ \bibinfo {author} {\bibfnamefont {C.~H.}\ \bibnamefont {Keitel}},\
  }\bibfield  {title} {\enquote {\bibinfo {title} {Implementing nonlinear
  compton scattering beyond the local-constant-field approximation},}\ }\href
  {\doibase 10.1103/PhysRevA.98.012134} {\bibfield  {journal} {\bibinfo
  {journal} {Phys. Rev. A}\ }\textbf {\bibinfo {volume} {98}},\ \bibinfo
  {pages} {012134} (\bibinfo {year} {2018})}\BibitemShut {NoStop}%
\bibitem [{\citenamefont {Ilderton}(2019)}]{Ilderton2019prd}%
  \BibitemOpen
  \bibfield  {author} {\bibinfo {author} {\bibfnamefont {A.}~\bibnamefont
  {Ilderton}},\ }\bibfield  {title} {\enquote {\bibinfo {title} {Note on the
  conjectured breakdown of qed perturbation theory in strong fields},}\ }\href
  {\doibase 10.1103/PhysRevD.99.085002} {\bibfield  {journal} {\bibinfo
  {journal} {Phys. Rev. D}\ }\textbf {\bibinfo {volume} {99}},\ \bibinfo
  {pages} {085002} (\bibinfo {year} {2019})}\BibitemShut {NoStop}%
\bibitem [{\citenamefont {Di~Piazza}\ \emph {et~al.}(2019)\citenamefont
  {Di~Piazza}, \citenamefont {Tamburini}, \citenamefont {Meuren},\ and\
  \citenamefont {Keitel}}]{piazza2019}%
  \BibitemOpen
  \bibfield  {author} {\bibinfo {author} {\bibfnamefont {A.}~\bibnamefont
  {Di~Piazza}}, \bibinfo {author} {\bibfnamefont {M.}~\bibnamefont
  {Tamburini}}, \bibinfo {author} {\bibfnamefont {S.}~\bibnamefont {Meuren}}, \
  and\ \bibinfo {author} {\bibfnamefont {C.~H.}\ \bibnamefont {Keitel}},\
  }\bibfield  {title} {\enquote {\bibinfo {title} {Improved
  local-constant-field approximation for strong-field qed codes},}\ }\href
  {\doibase 10.1103/PhysRevA.99.022125} {\bibfield  {journal} {\bibinfo
  {journal} {Phys. Rev. A}\ }\textbf {\bibinfo {volume} {99}},\ \bibinfo
  {pages} {022125} (\bibinfo {year} {2019})}\BibitemShut {NoStop}%
\bibitem [{sup()}]{supplemental}%
  \BibitemOpen
  \href@noop {} {\ }\bibinfo {note} {See Supplemental Material for details on
  the employed laser fields, the applied simulation methods, and the simulation
  results for other laser or $\gamma$ photon parameters.}\BibitemShut {Stop}%
\bibitem [{\citenamefont {Thomas}(1926)}]{Thomas_1926}%
  \BibitemOpen
  \bibfield  {author} {\bibinfo {author} {\bibfnamefont {L.~H.}\ \bibnamefont
  {Thomas}},\ }\bibfield  {title} {\enquote {\bibinfo {title} {The motion of
  the spinning electron},}\ }\href@noop {} {\bibfield  {journal} {\bibinfo
  {journal} {Nature (London)}\ }\textbf {\bibinfo {volume} {117}},\ \bibinfo
  {pages} {514} (\bibinfo {year} {1926})}\BibitemShut {NoStop}%
\bibitem [{\citenamefont {Thomas}(1927)}]{Thomas_1927}%
  \BibitemOpen
  \bibfield  {author} {\bibinfo {author} {\bibfnamefont {L.~H.}\ \bibnamefont
  {Thomas}},\ }\bibfield  {title} {\enquote {\bibinfo {title} {The kinematics
  of an electron with an axis},}\ }\href@noop {} {\bibfield  {journal}
  {\bibinfo  {journal} {Philos. Mag.}\ }\textbf {\bibinfo {volume} {3}},\
  \bibinfo {pages} {1--22} (\bibinfo {year} {1927})}\BibitemShut {NoStop}%
\bibitem [{\citenamefont {Bargmann}\ \emph {et~al.}(1959)\citenamefont
  {Bargmann}, \citenamefont {Michel},\ and\ \citenamefont
  {Telegdi}}]{Bargmann_1959}%
  \BibitemOpen
  \bibfield  {author} {\bibinfo {author} {\bibfnamefont {V.}~\bibnamefont
  {Bargmann}}, \bibinfo {author} {\bibfnamefont {Louis}\ \bibnamefont
  {Michel}}, \ and\ \bibinfo {author} {\bibfnamefont {V.~L.}\ \bibnamefont
  {Telegdi}},\ }\bibfield  {title} {\enquote {\bibinfo {title} {Precession of
  the polarization of particles moving in a homogeneous electromagnetic
  field},}\ }\href {\doibase 10.1103/PhysRevLett.2.435} {\bibfield  {journal}
  {\bibinfo  {journal} {Phys. Rev. Lett.}\ }\textbf {\bibinfo {volume} {2}},\
  \bibinfo {pages} {435--436} (\bibinfo {year} {1959})}\BibitemShut {NoStop}%
\bibitem [{\citenamefont {Salamin}\ and\ \citenamefont
  {Keitel}(2002)}]{Salamin2002}%
  \BibitemOpen
  \bibfield  {author} {\bibinfo {author} {\bibfnamefont {Yousef~I.}\
  \bibnamefont {Salamin}}\ and\ \bibinfo {author} {\bibfnamefont
  {Christoph~H.}\ \bibnamefont {Keitel}},\ }\bibfield  {title} {\enquote
  {\bibinfo {title} {Electron acceleration by a tightly focused laser beam},}\
  }\href {\doibase 10.1103/PhysRevLett.88.095005} {\bibfield  {journal}
  {\bibinfo  {journal} {Phys. Rev. Lett.}\ }\textbf {\bibinfo {volume} {88}},\
  \bibinfo {pages} {095005} (\bibinfo {year} {2002})}\BibitemShut {NoStop}%
\bibitem [{\citenamefont {Galow}\ \emph {et~al.}(2011)\citenamefont {Galow},
  \citenamefont {Salamin}, \citenamefont {Liseykina}, \citenamefont {Harman},\
  and\ \citenamefont {Keitel}}]{Galow_2011}%
  \BibitemOpen
  \bibfield  {author} {\bibinfo {author} {\bibfnamefont {Benjamin~J.}\
  \bibnamefont {Galow}}, \bibinfo {author} {\bibfnamefont {Yousef~I.}\
  \bibnamefont {Salamin}}, \bibinfo {author} {\bibfnamefont {Tatyana~V.}\
  \bibnamefont {Liseykina}}, \bibinfo {author} {\bibfnamefont {Zolt\'an}\
  \bibnamefont {Harman}}, \ and\ \bibinfo {author} {\bibfnamefont
  {Christoph~H.}\ \bibnamefont {Keitel}},\ }\bibfield  {title} {\enquote
  {\bibinfo {title} {Dense monoenergetic proton beams from chirped laser-plasma
  interaction},}\ }\href {\doibase 10.1103/PhysRevLett.107.185002} {\bibfield
  {journal} {\bibinfo  {journal} {Phys. Rev. Lett.}\ }\textbf {\bibinfo
  {volume} {107}},\ \bibinfo {pages} {185002} (\bibinfo {year}
  {2011})}\BibitemShut {NoStop}%
\bibitem [{\citenamefont {Baynham}\ \emph {et~al.}(2011)\citenamefont
  {Baynham}, \citenamefont {Ivanyushenkov}, \citenamefont {Clarke},
  \citenamefont {Brummitt}, \citenamefont {Bayliss}, \citenamefont {Bradshaw},
  \citenamefont {Scott}, \citenamefont {Rochford}, \citenamefont {Carr},
  \citenamefont {Burton}, \citenamefont {Taylor},\ and\ \citenamefont
  {Lintern}}]{Baynham_2011}%
  \BibitemOpen
  \bibfield  {author} {\bibinfo {author} {\bibfnamefont {D.~E.}\ \bibnamefont
  {Baynham}}, \bibinfo {author} {\bibfnamefont {Y.}~\bibnamefont
  {Ivanyushenkov}}, \bibinfo {author} {\bibfnamefont {J.~A.}\ \bibnamefont
  {Clarke}}, \bibinfo {author} {\bibfnamefont {A.}~\bibnamefont {Brummitt}},
  \bibinfo {author} {\bibfnamefont {V.}~\bibnamefont {Bayliss}}, \bibinfo
  {author} {\bibfnamefont {T.}~\bibnamefont {Bradshaw}}, \bibinfo {author}
  {\bibfnamefont {D.~J.}\ \bibnamefont {Scott}}, \bibinfo {author}
  {\bibfnamefont {J.}~\bibnamefont {Rochford}}, \bibinfo {author}
  {\bibfnamefont {S.}~\bibnamefont {Carr}}, \bibinfo {author} {\bibfnamefont
  {G.}~\bibnamefont {Burton}}, \bibinfo {author} {\bibfnamefont
  {O.}~\bibnamefont {Taylor}}, \ and\ \bibinfo {author} {\bibfnamefont
  {A.}~\bibnamefont {Lintern}},\ }\bibfield  {title} {\enquote {\bibinfo
  {title} {{Demonstration of a High-Field Short-Period Superconducting Helical
  Undulator Suitable for Future TeV-Scale Linear Collider Positron Sources}},}\
  }\href@noop {} {\bibfield  {journal} {\bibinfo  {journal} {Phys. Rev. Lett.}\
  }\textbf {\bibinfo {volume} {107}},\ \bibinfo {pages} {1--5} (\bibinfo {year}
  {2011})}\BibitemShut {NoStop}%
\bibitem [{\citenamefont {Abbott~{\it et al.}}(2016)}]{Abbott_2016}%
  \BibitemOpen
  \bibfield  {author} {\bibinfo {author} {\bibfnamefont {D.}~\bibnamefont
  {Abbott~{\it et al.}}} (\bibinfo {collaboration} {PEPPo Collaboration}),\
  }\bibfield  {title} {\enquote {\bibinfo {title} {Production of highly
  polarized positrons using polarized electrons at mev energies},}\ }\href@noop
  {} {\bibfield  {journal} {\bibinfo  {journal} {Phys. Rev. Lett.}\ }\textbf
  {\bibinfo {volume} {116}},\ \bibinfo {pages} {214801} (\bibinfo {year}
  {2016})}\BibitemShut {NoStop}%
\bibitem [{\citenamefont {King}\ and\ \citenamefont {Tang}(2020)}]{King_2020}%
  \BibitemOpen
  \bibfield  {author} {\bibinfo {author} {\bibfnamefont {B.}~\bibnamefont
  {King}}\ and\ \bibinfo {author} {\bibfnamefont {S.}~\bibnamefont {Tang}},\
  }\bibfield  {title} {\enquote {\bibinfo {title} {Nonlinear compton scattering
  of polarized photons in plane-wave backgrounds},}\ }\href {\doibase
  10.1103/PhysRevA.102.022809} {\bibfield  {journal} {\bibinfo  {journal}
  {Phys. Rev. A}\ }\textbf {\bibinfo {volume} {102}},\ \bibinfo {pages}
  {022809} (\bibinfo {year} {2020})}\BibitemShut {NoStop}%
\bibitem [{\citenamefont {Tang}\ \emph {et~al.}(2020)\citenamefont {Tang},
  \citenamefont {King},\ and\ \citenamefont {Hu}}]{Tang_2020}%
  \BibitemOpen
  \bibfield  {author} {\bibinfo {author} {\bibfnamefont {S.}~\bibnamefont
  {Tang}}, \bibinfo {author} {\bibfnamefont {B.}~\bibnamefont {King}}, \ and\
  \bibinfo {author} {\bibfnamefont {H.}~\bibnamefont {Hu}},\ }\bibfield
  {title} {\enquote {\bibinfo {title} {{Highly polarised gamma photons from
  electron-laser collisions}},}\ }\href@noop {} {\bibfield  {journal} {\bibinfo
   {journal} {Phys. Lett. B}\ }\textbf {\bibinfo {volume} {809}} (\bibinfo
  {year} {2020})}\BibitemShut {NoStop}%
\bibitem [{\citenamefont {Li}\ \emph {et~al.}(2020{\natexlab{b}})\citenamefont
  {Li}, \citenamefont {Shaisultanov}, \citenamefont {Chen}, \citenamefont
  {Wan}, \citenamefont {Hatsagortsyan}, \citenamefont {Keitel},\ and\
  \citenamefont {Li}}]{Ligammaray_2019}%
  \BibitemOpen
  \bibfield  {author} {\bibinfo {author} {\bibfnamefont {Yan-Fei}\ \bibnamefont
  {Li}}, \bibinfo {author} {\bibfnamefont {Rashid}\ \bibnamefont
  {Shaisultanov}}, \bibinfo {author} {\bibfnamefont {Yue-Yue}\ \bibnamefont
  {Chen}}, \bibinfo {author} {\bibfnamefont {Feng}\ \bibnamefont {Wan}},
  \bibinfo {author} {\bibfnamefont {Karen~Z.}\ \bibnamefont {Hatsagortsyan}},
  \bibinfo {author} {\bibfnamefont {Christoph~H.}\ \bibnamefont {Keitel}}, \
  and\ \bibinfo {author} {\bibfnamefont {Jian-Xing}\ \bibnamefont {Li}},\
  }\bibfield  {title} {\enquote {\bibinfo {title} {Polarized ultrashort
  brilliant multi-gev $\ensuremath{\gamma}$ rays via single-shot laser-electron
  interaction},}\ }\href {\doibase 10.1103/PhysRevLett.124.014801} {\bibfield
  {journal} {\bibinfo  {journal} {Phys. Rev. Lett.}\ }\textbf {\bibinfo
  {volume} {124}},\ \bibinfo {pages} {014801} (\bibinfo {year}
  {2020}{\natexlab{b}})}\BibitemShut {NoStop}%
\bibitem [{\citenamefont {Artru}\ \emph {et~al.}(2008)\citenamefont {Artru},
  \citenamefont {Chehab}, \citenamefont {Chevallier}, \citenamefont
  {Strakhovenko}, \citenamefont {Variola},\ and\ \citenamefont
  {Vivoli}}]{Artru_2008}%
  \BibitemOpen
  \bibfield  {author} {\bibinfo {author} {\bibfnamefont {X.}~\bibnamefont
  {Artru}}, \bibinfo {author} {\bibfnamefont {R.}~\bibnamefont {Chehab}},
  \bibinfo {author} {\bibfnamefont {M.}~\bibnamefont {Chevallier}}, \bibinfo
  {author} {\bibfnamefont {V.M.}\ \bibnamefont {Strakhovenko}}, \bibinfo
  {author} {\bibfnamefont {A.}~\bibnamefont {Variola}}, \ and\ \bibinfo
  {author} {\bibfnamefont {A.}~\bibnamefont {Vivoli}},\ }\bibfield  {title}
  {\enquote {\bibinfo {title} {Polarized and unpolarized positron sources for
  electron-positron colliders},}\ }\href {\doibase
  https://doi.org/10.1016/j.nimb.2008.02.086} {\bibfield  {journal} {\bibinfo
  {journal} {Nucl. Instrum. Methods Phys. Res., Sect. B}\ }\textbf {\bibinfo
  {volume} {266}},\ \bibinfo {pages} {3868 -- 3875} (\bibinfo {year}
  {2008})}\BibitemShut {NoStop}%
\bibitem [{\citenamefont {Leemans}\ and\ \citenamefont
  {Esarey}(2009)}]{Leemans2009}%
  \BibitemOpen
  \bibfield  {author} {\bibinfo {author} {\bibfnamefont {Wim}\ \bibnamefont
  {Leemans}}\ and\ \bibinfo {author} {\bibfnamefont {Eric}\ \bibnamefont
  {Esarey}},\ }\bibfield  {title} {\enquote {\bibinfo {title} {{Laser-driven
  plasma-wave electron accelerators}},}\ }\href {\doibase 10.1063/1.3099645}
  {\bibfield  {journal} {\bibinfo  {journal} {Phys. Today}\ }\textbf {\bibinfo
  {volume} {62}},\ \bibinfo {pages} {44--49} (\bibinfo {year}
  {2009})}\BibitemShut {NoStop}%
\bibitem [{\citenamefont {Luo}\ \emph {et~al.}(2018)\citenamefont {Luo},
  \citenamefont {Chen}, \citenamefont {Wu}, \citenamefont {Weng}, \citenamefont
  {Sheng}, \citenamefont {Schroeder}, \citenamefont {Jaroszynski},
  \citenamefont {Esarey}, \citenamefont {Leemans}, \citenamefont {Mori},\ and\
  \citenamefont {Zhang}}]{Luo2018}%
  \BibitemOpen
  \bibfield  {author} {\bibinfo {author} {\bibfnamefont {J.}~\bibnamefont
  {Luo}}, \bibinfo {author} {\bibfnamefont {M.}~\bibnamefont {Chen}}, \bibinfo
  {author} {\bibfnamefont {W.~Y.}\ \bibnamefont {Wu}}, \bibinfo {author}
  {\bibfnamefont {S.~M.}\ \bibnamefont {Weng}}, \bibinfo {author}
  {\bibfnamefont {Z.~M.}\ \bibnamefont {Sheng}}, \bibinfo {author}
  {\bibfnamefont {C.~B.}\ \bibnamefont {Schroeder}}, \bibinfo {author}
  {\bibfnamefont {D.~A.}\ \bibnamefont {Jaroszynski}}, \bibinfo {author}
  {\bibfnamefont {E.}~\bibnamefont {Esarey}}, \bibinfo {author} {\bibfnamefont
  {W.~P.}\ \bibnamefont {Leemans}}, \bibinfo {author} {\bibfnamefont {W.~B.}\
  \bibnamefont {Mori}}, \ and\ \bibinfo {author} {\bibfnamefont
  {J.}~\bibnamefont {Zhang}},\ }\bibfield  {title} {\enquote {\bibinfo {title}
  {Multistage coupling of laser-wakefield accelerators with curved plasma
  channels},}\ }\href {\doibase 10.1103/PhysRevLett.120.154801} {\bibfield
  {journal} {\bibinfo  {journal} {Phys. Rev. Lett.}\ }\textbf {\bibinfo
  {volume} {120}},\ \bibinfo {pages} {154801} (\bibinfo {year}
  {2018})}\BibitemShut {NoStop}%
\bibitem [{\citenamefont {Ananthanarayan}\ and\ \citenamefont
  {Rindani}(2004{\natexlab{b}})}]{Ananthanarayan_2004}%
  \BibitemOpen
  \bibfield  {author} {\bibinfo {author} {\bibfnamefont {B.}~\bibnamefont
  {Ananthanarayan}}\ and\ \bibinfo {author} {\bibfnamefont {Saurabh~D.}\
  \bibnamefont {Rindani}},\ }\bibfield  {title} {\enquote {\bibinfo {title} {Cp
  violation at a linear collider with transverse polarization},}\ }\href
  {\doibase 10.1103/PhysRevD.70.036005} {\bibfield  {journal} {\bibinfo
  {journal} {Phys. Rev. D}\ }\textbf {\bibinfo {volume} {70}},\ \bibinfo
  {pages} {036005} (\bibinfo {year} {2004}{\natexlab{b}})}\BibitemShut
  {NoStop}%
\end{thebibliography}%

\end{document}